\def\ARXIV{1} 

\documentclass[a4paper,fleqn]{cas-dc}

\usepackage{amsmath,amssymb,amsfonts}
\usepackage{graphicx}
\usepackage{booktabs}
\usepackage{array}
\usepackage{multirow}
\usepackage{xcolor}
\usepackage[numbers,sort&compress]{natbib}
\usepackage{url}

\usepackage{tabularx}
\usepackage{makecell}
\usepackage{adjustbox}
\usepackage{placeins}
\usepackage{capt-of}
\newcolumntype{Y}{>{\raggedright\arraybackslash}X}

\usepackage{hyphenat}
\ifnum\ARXIV=1
    \usepackage{orcidlink}

    \ExplSyntaxOn
    \cs_gset:Npn \__first_footerline:
      {
        \group_begin:
        \small \sffamily
        \__short_authors:
        \quad arXiv~preprint
        \group_end:
      }
    \ExplSyntaxOff
\fi

\begin{document}

\let\WriteBookmarks\relax
\def\floatpagepagefraction{1}
\def\textpagefraction{.001}

\shorttitle{Incentive Alignment for Electricity Data Assets in China}
\shortauthors{Zhang et al.}

\title[mode=title]{Incentive Alignment for Electricity Data Assets in China: A Blockchain-Enabled Contribution-Credit Framework}

\ifnum\ARXIV=1

    \author[1,4]{Jiazheng Zhang}[orcid=0000-0001-5707-6012]
    \ead{zjz@seu.edu.cn}
    
    \author[2]{Shouwei Li}[orcid=0000-0003-1534-1628]
    \ead{lishouwei@seu.edu.cn}
    
    \author[3]{Yutian Deng}
    \ead{dengyutian@epri.sgcc.com.cn}
    
    \author[4]{Wei Wei}[orcid=0000-0002-8361-0941]
    \ead{wei.wei@bristol.ac.uk}
    
    \author[4]{John Cartlidge}[orcid=0000-0002-3143-6355]
    \ead{john.cartlidge@bristol.ac.uk}
\else
    \author[1,4]{Jiazheng Zhang}[orcid=0000-0001-5707-6012]
    \cormark[1]
    \ead{zjz@seu.edu.cn}
    
    \author[2]{Shouwei Li}[orcid=0000-0003-1534-1628]
    \cormark[1]
    \ead{lishouwei@seu.edu.cn}
    
    \author[3]{Yutian Deng}
    \ead{dengyutian@epri.sgcc.com.cn}
    
    \author[4]{Wei Wei}[orcid=0000-0002-8361-0941]
    \ead{wei.wei@bristol.ac.uk}
    
    \author[4]{John Cartlidge}[orcid=0000-0002-3143-6355]
    \ead{john.cartlidge@bristol.ac.uk}
\fi

\affiliation[1]{
    organization={School of Cyber Science and Engineering, Southeast University},
    city={Nanjing},
    postcode={211189},
    country={China}
}

\affiliation[2]{
    organization={School of Economics and Management, Southeast University},
    city={Nanjing},
    postcode={211189},
    country={China}
}

\affiliation[3]{
    organization={China Electric Power Research Institute Co., Ltd.},
    city={Nanjing},
    postcode={210009},
    country={China}
}

\affiliation[4]{
    organization={School of Engineering Mathematics and Technology, University of Bristol},
    city={Bristol},
    postcode={BS8 1QU},
    country={United Kingdom}
}

\ifnum\ARXIV=1
\else
    \cortext[cor1]{Corresponding authors}
\fi

\begin{abstract}
Modern electricity markets increasingly rely on reported operational contributions to allocate incentives, recognise cooperation, and coordinate activities across multiple organisations. However, supporting evidence for many reported contributions becomes available only after subsequent metering, dispatch, settlement, or auditing processes, making it difficult to reward genuine contributions while discouraging strategic over-reporting. Existing approaches largely address data sharing, performance verification, blockchain record management, or incentive allocation in isolation, providing limited support for integrating delayed evidence into a unified incentive mechanism. This paper proposes a blockchain-enabled contribution-credit framework for truthful reporting under delayed evidence. The framework links authenticated evidence to contribution recognition through dynamic credit updating, escrow, report-linked penalties, and automated incentive execution. Developed using the Chinese electricity market as the primary setting, it is demonstrated through two representative coordination scenarios, with adaptations illustrating its applicability to UK and US electricity-market structures. A controlled heterogeneous-agent simulation evaluates the behavioural effects of the proposed mechanism, while security analysis, public-data workflow analysis, and blockchain implementation experiments assess its practical feasibility. The results show that integrating delayed evidence with dynamic credit, escrow, and report-linked penalties promotes truthful reporting while providing a transparent and auditable process for translating verified operational contributions into incentive allocation. The proposed framework offers a practical mechanism for improving accountability, coordination, and incentive compatibility in collaborative electricity markets.
\end{abstract}


\maketitle

\section{Introduction}
\noindent
Modern electricity coordination increasingly relies on
reported operational contributions to support incentive
decisions. These reports and their supporting records
constitute electricity data assets when they are governed
and reused for contribution recognition and responsibility
allocation. Their value depends on whether the information
can support reliable incentive decisions. However, a
contribution report may be submitted before the evidence
required to verify it becomes available. This delay creates
an opportunity for strategic over-reporting, particularly
when provisional rewards can be obtained before the
corresponding performance has been independently verified.

The central problem is therefore how to encourage truthful
reporting when verification is delayed. Existing research
has examined electricity-data governance, performance
incentives, and blockchain-enabled coordination
\cite{gough2022privacy,mitra2024smartmeteranalytics,
wasa2022contract,alasseur2020principalagent,
egunjobi2024blockchain,hassan2023oracle}.
However, limited attention has been paid to how an original
contribution report can be linked to evidence that becomes
available later and how the verification outcome can affect
current and future incentives. Section~2 reviews the
relevant literature and summarises this gap.

This paper addresses the problem through a blockchain-enabled
contribution-credit framework for reporting under delayed
verification. The framework links original reports to
authenticated evidence and converts the verification outcome
into enforceable incentive consequences. Dynamic credit
records reporting reliability across periods, while escrow
and report-linked penalties address the incentives created
by provisional rewards. The permissioned blockchain provides
an auditable coordination layer for preserving these
relationships and executing the corresponding rules.

The Chinese electricity system provides the primary setting
for developing the mechanism because its hierarchical
governance structure supports interprovincial coordination
and cross-organisational mutual aid
\cite{iea2023china,yan2024crossprovincial}.
The framework is instantiated in two representative scenarios:
intra-organisational interprovincial coordination and
cross-organisational mutual aid. It can also be extended to
the UK and US markets, as shown in Appendices~A and~B.

The paper makes three main contributions. First, it develops
a verification-aware contribution-credit mechanism and derives
a sufficient condition under which reporting beyond a
predefined tolerance cannot improve expected utility. This
identifies how delayed verification and its incentive
consequences affect strategic reporting. Second, the framework
is applied to two Chinese coordination scenarios with different
institutional incentive outputs. The UK and US extensions
further illustrate how the shared mechanism can be adapted
to different electricity-market arrangements. Third, the paper evaluates the efficiency implications of the mechanism in three bounded dimensions. Behavioural-resource efficiency is assessed through action compliance, report inflation, coordination effort, authenticated performance, participant utility, and transfer-neutral system surplus. Coordination-process efficiency is examined through the rule-level linkage of report registration, evidence verification, credit updating, and incentive execution. Blockchain execution efficiency is quantified through throughput, latency, success rate, and gas consumption in Ganache and Hyperledger Fabric. This separation identifies both the incentive gains delivered by delayed verification and the computational cost required to implement them.

The remainder of the paper is organised as follows.
Section~2 reviews the related literature. Section~3 develops
the China-centred mechanism, and Section~4 evaluates its
behavioural implications. Section~5 presents the security
analysis and implementation evaluation. Section~6 concludes
the paper. The appendices provide the UK and US extensions,
illustrative calculations, and additional public-data context.

\section{Related Work}
\label{sec:related-work}
\noindent
Research relevant to truthful reporting under delayed evidence
can be organised into four streams. These studies provide the
foundations for electricity-data coordination and incentive
design, but they address different parts of the reporting
problem. Table~\ref{tab:literature-gaps} summarises their
contributions and the remaining gaps relevant to this study.

The first stream examines how electricity data support shared
coordination. Smart-meter and electricity-data-governance
studies examine privacy protection and institutional data
management \cite{gough2022privacy,mitra2024smartmeteranalytics}.
Research on data as an economic asset explains how governance
and access rights affect the value of data, including its
non-rival use \cite{veldkamp2023data,jones2020data}.
Demand-response research further shows that participation
depends on measurable delivery and the institutional
arrangements through which flexibility is rewarded
\cite{cabot2024demand,morales2022demand}. These studies
establish the importance of governable data, but give limited
attention to how a contribution reported before verification
should affect subsequent incentives.

The second stream examines electricity-market performance and
institutional coordination. Research on renewable uncertainty
highlights the importance of reliable delivery and deviation
evidence \cite{zhu2025marketdesign,zhao2024weather}. Chinese
studies examine market reform and interprovincial coordination,
including the integration of renewable electricity
\cite{iea2023china,yan2024crossprovincial,xiang2023china,
yang2024greenpower,wang2023uncertainty}. Capacity-market
research considers resource adequacy and the consequences of
non-delivery, providing a foundation for the UK extension
\cite{tao2024capacity,desnz2024capacity}. Studies relevant to
the US extension examine data-centre demand and flexibility
\cite{masanet2020datacenter,iea2025energyai,colangelo2026ai}.
Research on electricity-attribute accounting also shows the
importance of traceable claims and temporal matching
\cite{bjorn2022certificates,xu2024system,riepin2024means,
scholta2025temporal}. Together, these studies establish the
institutional context of the proposed mechanism. However, the
connection between an earlier contribution report and its later
incentive consequences remains insufficiently developed across
these settings.

The third stream examines blockchain-enabled recording and
automated execution. Energy-blockchain studies investigate
trusted records and smart-contract applications in electricity
systems \cite{egunjobi2024blockchain,he2024review,
mazrae2025transactive,kirli2022smartcontracts,
junaidi2023blockchaindr}. These technologies can preserve an
auditable record of submitted information and execute
predefined rules. However, oracle research shows that on-chain
execution still depends on authenticated off-chain inputs
\cite{hassan2023oracle}. Blockchain recording therefore
provides an important technical foundation, but it does not by
itself change the private benefit of an inflated report. A
further mechanism is needed to link later verification to
enforceable incentive consequences.

The fourth stream examines incentive compatibility under
strategic reporting. Peer-prediction mechanisms study how
informative reports can be elicited when the ground truth is
not directly observable
\cite{miller2005peerprediction,zhao2024peerprediction}.
Contract-theoretic studies of power-grid moral hazard and
capacity remuneration examine incentive design under private
information and risk allocation
\cite{wasa2022contract,alasseur2020principalagent}.
Agent-based market models and incentive-based demand-response
simulations provide complementary methods for evaluating
heterogeneous responses to alternative rules
\cite{weidlich2008agent,barreto2014demandresponse}. These
studies provide the closest foundations for the present
analysis. Nevertheless, limited attention has been paid to the
joint effect of delayed authenticated evidence and persistent
credit on reporting incentives in electricity-data coordination.

\begin{table*}[!t]
\centering
\caption{Research Streams and Gaps Relevant to the Proposed Framework}
\label{tab:literature-gaps}
\scriptsize
\setlength{\tabcolsep}{5pt}
\renewcommand{\arraystretch}{1.12}
\begin{tabular}{@{}
>{\raggedright\arraybackslash}p{0.18\textwidth}
>{\raggedright\arraybackslash}p{0.38\textwidth}
>{\raggedright\arraybackslash}p{0.37\textwidth}@{}}
\toprule
Research stream
& Established contribution
& Remaining gap \\
\midrule

Electricity-data governance and sharing
&
Establishes the role of governable data and measurable
participation in electricity coordination.
&
Limited attention to the link between a report submitted
before verification and its later incentive consequences. \\

\addlinespace

Electricity-market performance and institutional coordination
&
Examines performance evidence and institutional arrangements
across electricity-market settings.
&
Limited integration of earlier reports and later evidence
into a common incentive-accountability framework. \\

\addlinespace

Blockchain-enabled recording and execution
&
Provides auditable records and rule-based execution, subject
to the reliability of off-chain evidence.
&
Record integrity alone does not address the private benefit
of inflated reporting. \\

\addlinespace

Strategic reporting and incentive compatibility
&
Develops incentive mechanisms and methods for evaluating
heterogeneous responses to alternative rules.
&
Limited joint treatment of delayed authenticated evidence
and persistent credit in the present setting. \\

\bottomrule
\end{tabular}
\end{table*}

Table~\ref{tab:literature-gaps} brings these research gaps
together. The central issue is the temporal link between an
original contribution report and the evidence that becomes
available later. This study addresses the issue through a
verification-aware contribution-credit framework that connects
delayed evidence to current and future incentive consequences.
The framework is developed for two Chinese coordination
scenarios in Section~3, while Section~4 evaluates its reporting
incentives through a controlled simulation.

\section{China-Centered Contribution-Credit Mechanism}
\begin{figure*}[!t]
    \centering
    \includegraphics[width=0.98\textwidth]{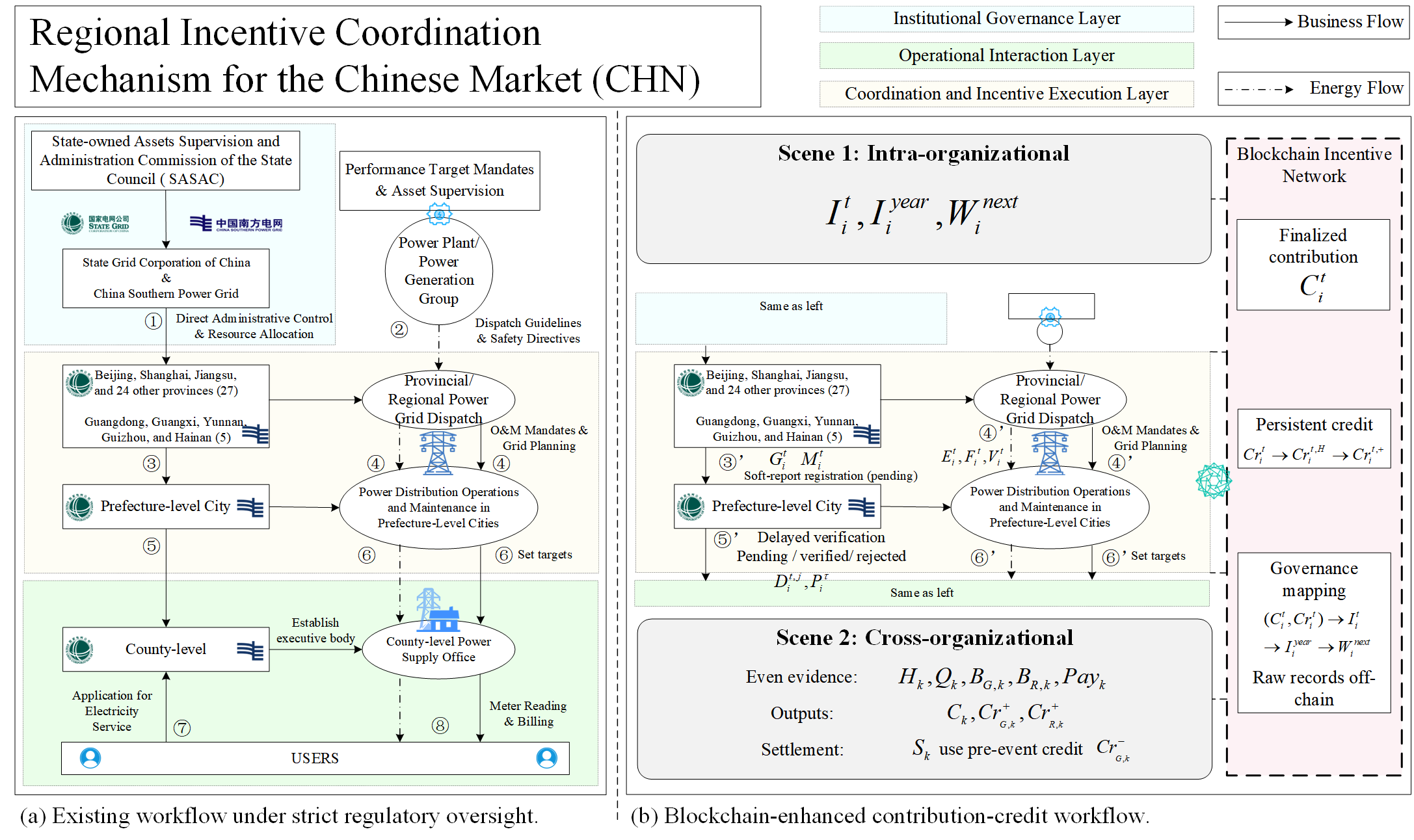}
    \caption{Layered comparison of the Chinese incentive-coordination workflow across the Institutional Governance, Coordination and Incentive Execution, and Operational Interaction layers. Subfigure (a) presents the existing hierarchical workflow, whereas subfigure (b) introduces the blockchain-enabled contribution-credit mechanism, supporting governance-oriented priority adjustment in Scene~1 and settlement-oriented accountability in Scene~2. For Scene~1, Steps~$3'$--$6'$ are the blockchain-enhanced counterparts of Steps~3--6.}
    \label{fig:china}
\end{figure*}

\noindent
This section develops the contribution-credit mechanism using China as the primary institutional setting. Its point of departure is that a participant may report a favourable operational contribution before all the evidence required to verify that report becomes available. As shown in Fig.~\ref{fig:china}, the mechanism connects the original report to subsequent operational evidence and then translates the verification outcome into contribution, credit, and incentive consequences.

Both panels of Fig.~\ref{fig:china} contain three functional layers: the Institutional Governance Layer, the Coordination and Incentive Execution Layer, and the Operational Interaction Layer. The upper layer defines governance, dispatch, and contractual constraints and receives the resulting incentive feedback. The middle layer registers reports, manages their verification states, and calculates contribution, credit, and incentive outputs. The lower layer supplies operational, metering, billing, and settlement evidence. The permissioned blockchain is located in the middle layer because its role is to preserve the relationship between an original report, the evidence that becomes available later, and the resulting incentive consequence.

Fig.~\ref{fig:china}(a) presents the existing workflow, while Fig.~\ref{fig:china}(b) preserves its institutional structure and introduces the contribution\hyp{}credit mechanism. Steps~$3'$--$6'$ correspond to Steps~3--6 but add report registration, evidence linking, contribution recognition, credit updating, and incentive feedback. A submitted report may remain pending until the relevant operational evidence becomes available; registration therefore does not imply immediate acceptance. Steps~1 and~2 continue to define governance and dispatch constraints, whereas Steps~7 and~8 provide demand-side, metering, and billing evidence. The prime notation identifies only the process stages modified by the mechanism.

The mechanism is instantiated in two Chinese coordination scenarios: intra-organisational interprovincial coordination and cross-organisational mutual aid. Their respective institutional settings and incentive mappings are detailed in Sections~3.1 and~3.2.

Because the China-centred mechanism is the only mechanism formalised in the main text, country subscripts are omitted in this section. Table~\ref{tab:symbols} summarises the core notation used in the two Chinese scenarios. Accordingly, Appendices~A and~B do not restate the common normalization, contribution, credit-updating, report-deviation, penalty, or incentive-compatibility equations. They apply Eqs.~(1)--(3) and Eqs.~(11)--(16) using UK- or US-indexed evidence sets, bounds, and parameters, and retain only the indicator interpretations, authenticated evidence channels, and upper-layer incentive mappings that differ across the two markets.

\begin{table}[!t]
\centering
\caption{Core Notation for the China-Centered Contribution-Credit Mechanism}
\label{tab:symbols}
\scriptsize
\setlength{\tabcolsep}{2pt}
\renewcommand{\arraystretch}{1.04}
\begin{tabularx}{\columnwidth}{p{0.27\columnwidth}Y}
\toprule
Symbol & Functional meaning \\
\midrule
$i$, $t$, $j$, $k$
& Entity, time-period, indicator, and mutual-aid-event indices. \\

$N$, $T$
& Participating entities and reporting periods. \\

$G_i^{t}$, $E_i^{t}$, $M_i^{t}$, $F_i^{t}$, $V_i^{t}$
& Green-coordination commitment, executed coordination, obligation-fulfilment improvement, fulfilment quality, and execution deviation. \\

$\mathcal{H}$, $\mathcal{S}$, $x_i^{t,j}$, $g_i^{t,j}$,
$\widehat{x}_{i}^{t,j}$, $z_{i}^{t+\ell_j,j}$
& Hard-evidence and soft-report sets, latent value, report inflation, submitted value, and delayed verified reference. \\


$C_i^{t}$, $Cr_i^{t}$, $Cr_i^{t,H}$,
$Cr_i^{\tau,-}$, $Cr_i^{\tau,+}$
& Finalized contribution, dynamic credit, post-hard-evidence credit, and credit before and after delayed report settlement. \\

$I_i^{t,0}$, $I_i^{t,\mathrm{exec}}$,
$I_i^{\mathrm{year}}$, $W_i^{\mathrm{next}}$
& Base and executed period incentives, annual incentive,
and next-period priority. \\

$H_k$, $Q_k$, $U_k$, $P_k$, $B_k$, $C_k$, $S_k^{0}$
& Mutual-aid volume, completion, urgency, price, event deviation,
event contribution, and base settlement. \\

$\mathrm{Adj}_{r,k}^{\tau}$,
$S_{G,k}^{\mathrm{recv}}$, $S_{R,k}^{\mathrm{pay}}$
& Due-period report adjustment, supporting-side receipt,
and requesting-side payment. \\

$Cr_{G,k}^{-}$, $Cr_{G,k}^{+}$, $Cr_{R,k}^{-}$,
$Cr_{R,k}^{+}$, $\widetilde{\mathrm{Pay}}_k$, $\mathrm{Pen}_k$
& Pre- and post-event bilateral credit, payment compliance, and contractual penalty. \\

$D_i^{t,j}$, $A_i^{t,j}$, $\Pi_i^t$
& Tolerance-adjusted report deviation, audit indicator, and realized report-linked penalty. \\

$\varepsilon_j$, $p_j$, $\pi_j$, $\ell_j$, $q_{E,j}$, $\delta$
& Tolerance, audit probability, penalty intensity, verification delay, escrow fraction, and behavioural discount factor. \\

$Q_j(g)$, $\overline D_j(g)$, $w_j$, $\bar b_i^{t,j}$,
$c_{i,j}$, $\eta$
& Deviation probability and mean, reward weight, private marginal gain, cost curvature, and shadow value of future credit. \\

$\alpha$, $\beta$, $\rho$, $\lambda$, $\kappa$
& Scene~1 scoring, credit, memory, and incentive-mapping parameters. \\

$\gamma$, $\zeta$, $\mu$
& Scene~2 event-contribution, bilateral-credit, and settlement-mapping parameters. \\
\bottomrule
\end{tabularx}
\end{table}

As shown in Fig.~\ref{fig:china}(a), Steps~1 and~2 define the governance and operational boundary conditions of the existing workflow. SASAC supervises State Grid Corporation of China and China Southern Power Grid, which transmit administrative requirements and resource-allocation priorities to provincial and regional entities through Step~1. Power-generation groups provide generation plans, availability information, and plant-side safety constraints to provincial and regional dispatch centres through Step~2. These steps constrain the coordination process but are not themselves treated as contribution-scoring targets.

Steps~3--6 represent the principal coordination chain. Step~3 transfers provincial coordination tasks to prefecture-level entities, while Step~4 communicates dispatch, operation-and-maintenance, and planning requirements to local execution units. Steps~5 and~6 translate these requirements into county-level targets and execution arrangements. Steps~7 and~8 provide service, metering, and billing records from the Operational Interaction Layer. Although this hierarchy supports administrative control and routine operation, incentive-relevant reports and their subsequent evidence may remain distributed across organisational levels. Consequently, a report submitted when a coordination task is assigned may not be consistently linked to the execution and fulfilment evidence that becomes available later.

Fig.~\ref{fig:china}(b) preserves the institutional hierarchy but inserts the blockchain incentive network into the Coordination and Incentive Execution Layer. For Scene~1, Step~$3'$ registers the reported green-coordination commitment $G_i^{t}$ and obligation-fulfilment improvement $M_i^{t}$. Step~$4'$ supplies the executed coordination volume $E_i^{t}$, fulfilment rate $F_i^{t}$, and associated operational evidence. Step~$5'$ records the original submission, its timestamp and verification conditions, and subsequently links the report to the authenticated reference when that evidence becomes available. Until then, a soft report remains pending and any report-based incentive remains provisional.

Step~$6'$ converts finalised evidence into the contribution score $C_i^{t}$ and the applicable dynamic-credit state. It also releases or recovers provisional rewards, applies any report-linked penalty, and maps the finalised outputs to period incentives and future priority adjustment. Scene~2 follows the same report--evidence--consequence sequence but uses the event-level variables and settlement rules defined below. The blockchain therefore preserves and executes the connection between reports and later evidence; it does not independently determine whether the underlying off-chain evidence is true.

\subsection{Scene 1: Intra-Organizational Incentives for Interprovincial Coordination}
\noindent
China's unified national electricity market increasingly relies on interprovincial coordination within a hierarchical grid-governance structure \cite{iea2023china,yang2024greenpower,yan2024crossprovincial}. Even under common corporate control, provincial entities may face different local costs and reliability constraints when coordinating interprovincial transfers, renewable-energy accommodation, and transmission-corridor use. Sending-side entities must balance electricity exports against local reliability, whereas receiving-side entities must integrate imported electricity within their own operational constraints. Administrative coordination requirements therefore do not by themselves make active cooperation individually rational.

The central problem is information asymmetry between headquarters-level governance and provincial-level operation. Reported green-coordination and obligation\hyp{}improvement indicators can be strategically inflated, while costly coordination effort may remain below the level required for system-wide efficiency. Executed coordination, fulfilment quality, and deviations may become observable only later or remain stored in separate operational systems. Scene~1 therefore links submitted indicators to subsequent evidence and converts verified performance into contribution scores, dynamic credit, and future governance priorities, creating joint incentives for tolerance-compliant reporting and sustained coordination.

\subsubsection{Auditable Data Inputs and Verification States}
\noindent
Scene~1 uses a permissioned consortium blockchain as an auditable coordination layer, not as a source of data truth. For each entity $i\in N$ and source period $t=1,\ldots,T$, the ledger records the submitted value, timestamp, digital signature, verification due date, and current status. A reference to the supporting evidence is added when that evidence becomes available. Detailed metering, dispatch, settlement, and audit records remain in authorised off-chain repositories and are supplied by the responsible institutions when verification is required.

The period-level indicators are $G_i^{t}$, $E_i^{t}$, $M_i^{t}$, $F_i^{t}$, and $V_i^{t}$. Here, $G_i^{t}$ denotes the green-coordination commitment, $E_i^{t}$ the metered coordination volume actually executed, $M_i^{t}$ the improvement in obligation fulfilment relative to an ex ante baseline, $F_i^{t}$ the fulfilled share of the assigned coordination task, and $V_i^{t}$ the deviation from the registered commitment or dispatch instruction. The initially submitted values of $G_i^{t}$ and $M_i^{t}$ are treated as soft reports because their favourable content cannot be fully authenticated at submission. By contrast, $E_i^{t}$, $F_i^{t}$, and $V_i^{t}$ enter the hard-evidence set once the corresponding operational records have been authenticated. In the delayed-verification model below, the initial value of a soft report is denoted generically by $\widehat{x}_i^{t,j}$ and its later authenticated reference by $z_i^{t+\ell_j,j}$. Only the accepted value is used to finalise the contribution attributed to source period $t$.

In Fig.~\ref{fig:china}(b), Step~$3'$ registers the coordination\hyp{}commitment and obligation-improvement reports, while Step~$4'$ supplies execution, fulfilment, and deviation evidence. Step $5'$ preserves the link between each original report and its supporting records and assigns a pending, verified, or rejected status. A pending soft report may enter a provisional incentive calculation, but the corresponding report-based reward remains subject to escrow and cannot increase persistent credit. Once the required evidence becomes available, the report is verified against the authenticated reference and is then accepted, adjusted, or penalised according to the rules defined in Section~\ref{subsec:shared-delayed-verification}.
To make the five indicators comparable, each submitted or authenticated value is normalised to $[0,1]$ using bounds fixed before the reporting period:
\begin{equation}
\begin{aligned}
\widetilde{X}_i^{t}
&=\operatorname{clip}_{[0,1]}
\left(\frac{X_i^{t}-L_{iX}^{t}}
{U_{iX}^{t}-L_{iX}^{t}}\right), \qquad \forall X\in\{G,E,M,F,V\}.
\end{aligned}
\label{eq:normalization}
\end{equation}

The lower and upper bounds, $L_{iX}^{t}$ and $U_{iX}^{t}$, are determined before reports are submitted using policy thresholds, assigned obligations, feasible capacity, or a predefined historical reference window, with $U_{iX}^{t}>L_{iX}^{t}$. The same bounds are applied to the initial soft report and its later authenticated reference, ensuring that any measured discrepancy is not created by a change in the normalisation rule. Fixing these bounds in advance also prevents an entity from influencing its normalised score through its current-period submission. The operator $\operatorname{clip}_{[0,1]}(\cdot)$ restricts the resulting value to the interval $[0,1]$.

\subsubsection{Contribution Scoring and Dynamic Credit Updating}
\noindent
After the required evidence for source period $t$ has been finalized, the contribution score of entity $i$ is
\begin{equation}
\begin{aligned}
C_i^{t}={}&\alpha_G\widetilde{G}_i^{t}
+\alpha_E\widetilde{E}_i^{t}+\alpha_M\widetilde{M}_i^{t}
+\alpha_F\widetilde{F}_i^{t}.
\end{aligned}
\label{eq:contribution_score}
\end{equation}

The weights are nonnegative, $\alpha_G,\alpha_E,\alpha_M,\alpha_F\geq0$, and satisfy $\alpha_G+\alpha_E+\alpha_M+\alpha_F=1$. Because all four inputs are normalized to $[0,1]$, $C_i^{t}\in[0,1]$. A component based on a soft report remains provisional until its due-date settlement; once accepted, it is assigned to its original source period. The coefficients reflect the relative importance of coordination commitment, verified execution, obligation improvement, and fulfilment performance.

Dynamic credit records execution and reporting reliability across periods. To keep hard evidence separate from delayed soft-report settlement, it is updated in two stages. Authenticated hard evidence in period $t$ produces
\begin{subequations}
\label{eq:credit_update}
\begin{align}
Cr_i^{t,H}={}&\operatorname{clip}_{[0,1]}\Bigl\{
Cr_i^{t}
+ (1-\rho)\bigl[
\beta_E(\widetilde E_i^{t}-Cr_i^{t}) \notag\\
&\qquad
+\beta_F(\widetilde F_i^{t}-Cr_i^{t})
\bigr]
-\beta_V\widetilde V_i^{t}
\Bigr\},
\label{eq:hard_credit_update}
\end{align}
while the soft reports from source periods $t$ that become due in period $\tau$ produce
\begin{align}
Cr_i^{\tau,+}={}&\operatorname{clip}_{[0,1]}\Biggl\{
Cr_i^{\tau,-}
+ (1-\rho)
\sum_{\substack{j\in\mathcal S\\ \tau\geq\ell_j}}
\beta_j A_i^{\tau-\ell_j,j} \notag\\
&\qquad\times
\bigl(\widetilde z_i^{\tau,j}-Cr_i^{\tau,-}\bigr)
-\beta_P\Pi_i^{\tau}
\Biggr\}.
\label{eq:soft_credit_update}
\end{align}
\end{subequations}
Here, $Cr_i^{t,H}$ is credit after the hard-evidence update, while $Cr_i^{\tau,-}$ and $Cr_i^{\tau,+}$ are the credit values immediately before and after settlement of the soft reports due in $\tau$. The normalized authenticated reference is $\widetilde z_i^{\tau,j}$. All credit weights are nonnegative, $0\leq\rho\leq1$, and the evidence weights are selected so that the pre-clipping mean-reversion coefficients are nonnegative. The operator $\operatorname{clip}_{[0,1]}(x)=\min\{1,\max\{0,x\}\}$ keeps credit within $[0,1]$.

The audit indicator $A_i^{t,j}$ and realized report\hyp{}linked pen\-alty $\Pi_i^{\tau}$ are defined in Section~\ref{subsec:shared-delayed-verification}. An unaudited soft report neither raises persistent credit nor incurs a report penalty. An audited and accepted report moves credit toward its authenticated reference; a detected deviation also incurs the nominal penalty through $\Pi_i^{\tau}$. Thus, the contribution score measures finalized source-period contribution, whereas dynamic credit accumulates authenticated execution and reporting reliability. These two outputs are combined only at the governance-oriented incentive-mapping stage.

\subsubsection{Governance-Oriented Incentive Mapping and Contract Execution}
\noindent
In Scene~1, contribution and credit outputs are mapped to
governance-oriented incentive scores rather than direct monetary
transfers. The base score and the score executed after settling any
report-linked penalties due in period $t$ are

\begin{subequations}
\label{eq:period_incentive}
\begin{align}
I_i^{t,0}
={}&\lambda_C C_i^{t}+\lambda_R Cr_i^{t,-},
\label{eq:period_incentive_base}\\
I_i^{t,\mathrm{exec}}
={}&\operatorname{clip}_{[0,1]}
\left(I_i^{t,0}-\Pi_i^{t}\right).
\label{eq:period_incentive_exec}
\end{align}
\end{subequations}

Here, $\lambda_C,\lambda_R\geq0$ and
$\lambda_C+\lambda_R=1$, and $Cr_i^{t,-}$ is the credit available
immediately before the due-report settlement.
Equation~\eqref{eq:period_incentive_exec} makes the direct consequence
of Eq.~\eqref{eq:realized-delayed-penalty-revised} explicit. The
nominal report penalty is deducted once from the governance-oriented
score executed in its due period. The same nominal penalty also enters
the credit update through $-\beta_P\Pi_i^t$ and therefore affects
future scores through credit. Escrow forfeiture is accounted for
separately and is not included in $\Pi_i^t$.

The annual incentive score is calculated as

\begin{equation}
I_i^{\mathrm{year}}
=\frac{1}{T}\sum_{t=1}^{T}I_i^{t,\mathrm{exec}}.
\label{eq:annual_incentive}
\end{equation}

Only finalized executed scores enter the annual calculation.
Any component based on a soft indicator remains provisional until
the corresponding delayed verification is completed.

Let $W_i^{0}$ denote the baseline priority weight of entity $i$, where $W_i^{0}\geq0$ and $\sum_{i\in N}W_i^{0}=1$. The next-period priority weight is updated as
\begin{equation}
\begin{aligned}
W_i^{\mathrm{next}} ={}&(1-\kappa)W_i^{0}+\kappa \frac{I_i^{\mathrm{year}}} {\sum_{j\in N}I_j^{\mathrm{year}}},
\end{aligned}
\label{eq:priority_update}
\end{equation}
where $0\leq\kappa\leq1$ controls the adjustment intensity. This formulation preserves $\sum_{i\in N}W_i^{\mathrm{next}}=1$ and prevents incentive adjustment from changing the total amount of priority available. If $\sum_{j\in N}I_j^{\mathrm{year}}=0$, the baseline allocation is retained, that is, $W_i^{\mathrm{next}}=W_i^{0}$.

Operationally, the smart contract checks digital signatures, evidence references, and verification states before applying Eqs.~\eqref{eq:period_incentive}--\eqref{eq:priority_update}. It then records the finalized incentive scores and proposed priority adjustments on the consortium ledger. In Fig.~\ref{fig:china}(b), the displayed $I_i^{t}$ denotes
$I_i^{t,\mathrm{exec}}$; this value, $I_i^{\mathrm{year}}$,
and $W_i^{\mathrm{next}}$ correspond to the governance-feedback
output of Step~$6'$. The resulting priority weight serves as an input to subsequent coordination and resource-allocation decisions; it does not replace operator approval or override dispatch and security constraints.

The two scenarios share the same mechanism-level sequence:
original report registration, evidence authentication, delayed
verification where necessary, contribution recognition, credit
updating, and rule-based execution. They differ only in their
institutional mappings. Scene~1 maps the resulting contribution
and credit values to governance-oriented priority adjustment,
whereas Scene~2 maps them to settlement and mutual-aid
accountability.

\subsection{Scene 2: Cross-Organizational Mutual-Aid Settlement and Accountability}
\noindent
Scene~2 considers mutual-aid events between the two major Chinese grid organizations, which operate under separate corporate and settlement boundaries. Under peak-load or emergency conditions, one organization may face temporary supply stress while the other can coordinate transferable support. Unlike Scene~1, such cooperation cannot be governed entirely through an internal administrative hierarchy. Its incentive outputs are therefore mapped to event-level settlement and bilateral credit rather than internal priority adjustment.

Existing arrangements already enable dispatch coordination and cross-network support. However, request registration, dispatch instructions, metering evidence, settlement records, and payment status may remain distributed across different systems. This fragmentation makes it difficult to link what was requested, what
was delivered, which party caused a deviation, and how the resulting responsibility should affect settlement and future cooperation. Scene~2 therefore establishes an auditable event record that connects registered obligations with verified execution and rule-based consequences.

As shown in Fig.~\ref{fig:china}(b), Scene~2 uses the same blockchain coordination layer as Scene~1 but applies a different upper-layer mapping. The figure highlights delivered support $H_k$, completion level $Q_k$, aggregate deviation $B_k$, and settlement output $S_k$. The urgency factor $U_k$, settlement price $P_k$, event contribution $C_k$, and bilateral credit states are defined below.

\subsubsection{Event Registration and Auditable Evidence}
\noindent
Each mutual-aid action is represented by an event $k$. The subscripts $G$ and $R$ denote the organizations occupying the supporting and requesting roles in that event, respectively; credit remains attached to the organization's ledger identity if these roles change in a later event.

Before execution, the authorized parties register the support obligation, response window, urgency classification $U_k$, and agreed settlement price $P_k$. The urgency classification is fixed by the authorized dispatch body rather than self-declared by either participant. After execution, authenticated dispatch and metering records determine the delivered electricity volume $H_k$ and completion level $Q_k$. The variables $B_{G,k}$ and $B_{R,k}$ record deviations attributable to the supporting and requesting sides, respectively. Specifically, $B_{G,k}$ captures deviation from the registered delivery profile, whereas $B_{R,k}$ captures deviation from the registered request or acceptance obligation. Payment and settlement records determine the requesting side's compliance indicator $\mathrm{Pay}_k$.

The consortium ledger stores the event identifier, timestamps, digital signatures, evidence references, and verification states, while detailed dispatch, metering, and payment records remain in authorized off-chain repositories. All performance indicators are normalized to $[0,1]$ using the ex ante normalization rule defined
above. The contractual price $P_k$ retains its original monetary unit for settlement calculation. In Fig.~\ref{fig:china}(b), the single variable $B_k$ visually represents the event's aggregate deviation, while the analytical model separates it into $B_{G,k}$ and $B_{R,k}$ for responsibility attribution.

\subsubsection{Event Contribution and Bilateral Credit Updating}
\noindent
The verified event contribution of the supporting organization is defined as
\begin{equation}
\begin{aligned}
C_k={}&\operatorname{clip}_{[0,1]}\Bigl(
\gamma_H\widetilde H_k+\gamma_Q\widetilde Q_k
+\gamma_U\widetilde U_k\widetilde Q_k
-\gamma_B\widetilde B_{G,k}
\Bigr).
\end{aligned}
\label{eq:event_contribution}
\end{equation}

Here, $\gamma_H,\gamma_Q,\gamma_U\geq0$, $\gamma_H+\gamma_Q+\gamma_U=1$, and $\gamma_B\geq0$. The interaction $\widetilde{U}_k\widetilde{Q}_k$ ensures that event urgency increases contribution only when the registered support is actually fulfilled. Urgency alone therefore cannot generate a positive contribution.

Let $Cr_{G,k}^{-}$ and $Cr_{G,k}^{+}$ denote the supporting organization's credit immediately before and after event $k$, respectively. Its credit is updated as
\begin{equation}
\begin{aligned}
Cr_{G,k}^{+} ={}&\operatorname{clip}_{[0,1]}\Bigl\{
Cr_{G,k}^{-} \\
&+(1-\rho)\Bigl[
\zeta_Q(\widetilde Q_k-Cr_{G,k}^{-}) \\
&\qquad
+\zeta_U(\widetilde U_k\widetilde Q_k-Cr_{G,k}^{-})
\Bigr] \\
&-\zeta_{BG}\widetilde B_{G,k}-\beta_P\Pi_{G,k}
\Bigr\}.
\end{aligned}
\label{eq:support_credit}
\end{equation}

\noindent Similarly, the requesting organization's credit is updated as
\begin{equation}
\begin{aligned}
Cr_{R,k}^{+} ={}&\operatorname{clip}_{[0,1]}\Bigl\{
Cr_{R,k}^{-} \\
&+(1-\rho)\bigl(
\widetilde{\mathrm{Pay}}_k-Cr_{R,k}^{-}
\bigr) \\
&-\zeta_{BR}\widetilde B_{R,k}-\beta_P\Pi_{R,k}
\Bigr\}.
\end{aligned}
\label{eq:request_credit}
\end{equation}

Here, $\zeta_Q,\zeta_U\geq0$ and $\zeta_Q+\zeta_U\leq1$, while $\zeta_{BG},\zeta_{BR}\geq0$ determine the effects of role-specific deviations. The event-linked penalties $\Pi_{G,k}$ and $\Pi_{R,k}$ are scaled by the same report-penalty credit weight $\beta_P$ used in Scene~1; each is zero when no report associated with that organisation becomes due for verification. These mean-reverting updates use the same hard-evidence, delayed-soft-evidence, and realized-penalty logic as Eq.~\eqref{eq:credit_update}, while retaining role-specific inputs.

These equations distinguish immediate event value from persistent bilateral reliability. The event contribution $C_k$ enters the settlement calculation for event $k$, whereas the post-event credit states $Cr_{G,k}^{+}$ and $Cr_{R,k}^{+}$ affect subsequent mutual-aid decisions. This separation prevents the same event performance from being counted twice in its own settlement.

\subsubsection{Settlement Mapping and Rule-Based Execution}
\noindent
In Scene~2, the verified event contribution and the supporting
organization's historical credit determine a base event settlement.
A report-linked penalty is then converted into a due-date monetary
adjustment for the participant occupying role $r\in\{G,R\}$.

\begin{subequations}
\label{eq:event_settlement}

\begin{equation}
S_k^{0}
=P_kH_k+\mu_C C_k+\mu_R Cr_{G,k}^{-}-Pen_k.
\label{eq:event_base_settlement}
\end{equation}

\begin{equation}
\mathrm{Adj}_{r,k}^{\tau}=\mu_{\Pi,r}\Pi_{r,k}^{\tau},
\qquad r\in\{G,R\}.
\label{eq:event_report_adjustment}
\end{equation}

\begin{equation}
\begin{aligned}
S_{G,k}^{\mathrm{recv}}
&=S_k^{0}-\mathrm{Adj}_{G,k}^{\tau}, \qquad S_{R,k}^{\mathrm{pay}}=S_k^{0}+\mathrm{Adj}_{R,k}^{\tau}.
\end{aligned}
\label{eq:event_net_settlement}
\end{equation}

\end{subequations}

Here, $P_k$ is the base unit price registered before execution,
and $H_k$ is the delivered electricity volume confirmed by
authenticated metering records. Because $C_k$ and $Cr_{G,k}^{-}$
are bounded within $[0,1]$, $\mu_C$ and $\mu_R$ represent the
maximum monetary premiums assigned to verified event contribution
and pre-event reliability, respectively. The coefficient
$\mu_{\Pi,r}\geq0$ converts the normalized report penalty into the
monetary unit used by the authorized settlement account.

A supporting-side report penalty reduces the amount received by the
supporting organization, while a requesting-side report penalty is
charged as an additional amount payable by the requesting organization.
The resulting adjustment is assigned to the authorized penalty or
coordination account and is not treated as an unregistered transfer
to the counterparty.

The term $Pen_k$ is the monetary penalty for a verified contractual
performance breach and is deducted in the base event settlement.
It is distinct from the delayed misreporting adjustment
$\mathrm{Adj}_{r,k}^{\tau}$. The latter is executed when the linked
report becomes verifiable at its due date $\tau$. If the base event
settlement remains open, the adjustment is netted in that settlement.
If the base settlement has already closed, it is posted as a separate
due-period debit without rewriting the finalized event record.

At the same due date, $\Pi_{r,k}^{\tau}$ also enters the corresponding
credit update through $-\beta_P\Pi_{r,k}^{\tau}$, thereby affecting
subsequent events. Thus, the direct settlement adjustment and the
persistent-credit reduction are separate consequences of the same
verified report deviation.

Operationally, the smart contract first checks the event identifier,
digital signatures, evidence references, and verification states.
Once the required dispatch and metering evidence has been authenticated,
the contract calculates $C_k$, $S_k^{0}$, and any due report adjustment
according to the registered rules. It then records the settlement
result and, where an authorized settlement interface is available,
issues the corresponding payment instruction.

In Fig.~\ref{fig:china}(b), $H_k$, $Q_k$, and $B_k$ represent the
principal execution evidence for Scene~2, while the displayed $S_k$
denotes the finalized settlement instruction produced by
Eq.~\eqref{eq:event_net_settlement}.

\subsection{Shared Delayed Verification and Report-Linked Penalties}
\label{subsec:shared-delayed-verification}
\noindent
The preceding contribution, credit, and settlement equations determine how accepted evidence is converted into incentive outcomes. Some favorable claims, however, cannot be authenticated when submitted. The mechanism therefore distinguishes a hard-evidence set $\mathcal H$ from a soft-report set $\mathcal S$. Authenticated metering, dispatch, settlement, payment, and operator-assigned records belong to $\mathcal H$. A participant's favorable commitment or improvement claim remains in $\mathcal S$ until an independent reference becomes available. This classification concerns evidence availability; it does not allow a participant to self-select the reference or baseline.

For each soft indicator $j\in\mathcal S$, the authorised governance body fixes the tolerance $\varepsilon_j$, audit probability $p_j$, penalty intensity $\pi_j$, verification delay $\ell_j$, escrow fraction $q_{E,j}$, and behavioural discount factor $\delta$ before reports are submitted. Let $x_i^{t,j}$ denote the latent value, $g_i^{t,j}\geq0$ upward report inflation, and $\widehat{x}_i^{t,j}=x_i^{t,j}+g_i^{t,j}$ the submitted report. At $t+\ell_j$, an authenticated reference is written as $z_i^{t+\ell_j,j}=x_i^{t,j}+\nu_{i,z}^{t,j}$, where $\nu_{i,z}^{t,j}$ represents reference-measurement error. The strategically relevant verified deviation is
\begin{equation}
\label{eq:report-deviation-revised}
D_i^{t,j}=\left[ \widehat{x}_i^{t,j}-z_i^{t+\ell_j,j}-\varepsilon_j \right]_+,
\end{equation}

where $[a]_+=\max\{a,0\}$. Indicators for which a lower value is favorable are reoriented before normalization, so that Eq.~\eqref{eq:report-deviation-revised} always measures an upward favorable claim.

The report enters the ledger as pending at $t$ and becomes due at $t+\ell_j$. Let $A_i^{t,j}\sim\mathrm{Bernoulli}(p_j)$ denote the audit indicator. A fraction $q_{E,j}$ of the report-based provisional incentive is held in escrow. At the due date, an unaudited report or an audited report with $D_i^{t,j}=0$ releases its escrow; an audited report with $D_i^{t,j}>0$ forfeits it and triggers the nominal penalty. Thus, the realized report-linked penalty processed in period $\tau$ is
\begin{equation}
\label{eq:realized-delayed-penalty-revised}
\Pi_i^{\tau}=
\sum_{\substack{j\in\mathcal S\\ \tau\geq\ell_j}}
A_i^{\tau-\ell_j,j}\pi_j D_i^{\tau-\ell_j,j}.
\end{equation}

The index $\tau-\ell_j$ links the consequence to the historical
report that has just become verifiable. At the due date, an audited
deviation has three separate consequences. First, the escrowed share
of the provisional report reward is forfeited. Second, the nominal
term $A_i^{t,j}\pi_jD_i^{t,j}$ is deducted directly from the incentive
or settlement account executed in the due period. Third, the same
nominal term enters the persistent-credit update through
$-\beta_P\Pi_i^{\tau}$ and therefore affects subsequent periods or
events.

An unaudited report or an audited report with $D_i^{t,j}=0$ triggers
neither the direct penalty nor the credit penalty. The discount factor
$\delta^{\ell_j}$ affects ex-ante valuation, not the amount executed
at the due date. Escrow forfeiture is not counted again in
$\Pi_i^{\tau}$.

In Scene~1, $\pi_j$ is calibrated in normalized incentive-score units,
and the direct deduction is implemented by
Eq.~\eqref{eq:period_incentive_exec}. In Scene~2,
$\Pi_{r,k}^{\tau}$ is converted to a monetary due-date adjustment by
Eq.~\eqref{eq:event_report_adjustment}. Its role-specific value also
enters the supporting- or requesting-side credit update. The
contractual-performance penalty $Pen_k$ remains separate from this
report-linked consequence.

To state the reporting incentive, define
\begin{equation}
\label{eq:general-deviation-moments}
Q_j(g)=\Pr\!\left(D_i^{t,j}>0\mid g\right),\qquad \overline D_j(g)=\mathbb E\!\left[D_i^{t,j}\mid g\right].
\end{equation}

Let $w_j$ be the reward weight of indicator $j$, $\bar b_i^{t,j}$ the net private marginal gain from report inflation, $c_{i,j}>0$ its cost curvature, $\eta$ the shadow value of future credit, and $\beta_P$ the report-penalty weight in the credit update. The report-related expected utility is
\begin{equation}
\label{eq:report-expected-utility-revised}
\begin{aligned}
U_{i,j}^{t,r}(g)={}& w_j(1-q_{E,j})(x_i^{t,j}+g) \\ &+\delta^{\ell_j}q_{E,j}w_j(x_i^{t,j}+g)
\big[1-p_jQ_j(g)\big] \\ &+\bar b_i^{t,j}g-\frac{c_{i,j}}{2}g^2 \\ &-\delta^{\ell_j}p_j\pi_j
(1+\eta\beta_P)\overline D_j(g).
\end{aligned}
\end{equation}

The first line is the immediately released report reward, and the
second is the expected discounted escrow release. In the final line,
the coefficient $1$ in $(1+\eta\beta_P)$ represents the direct
due-period deduction of the realized report penalty, while
$\eta\beta_P$ represents the shadow value of the associated
persistent-credit reduction. Escrow forfeiture is already reflected
in the release probability in the second line.
Equation~\eqref{eq:report-expected-utility-revised} is therefore
aligned with the two executed consequences of
Eq.~\eqref{eq:realized-delayed-penalty-revised}, and
Eq.~\eqref{eq:exact-reference-compliance-condition} retains the same
interpretation.

\noindent\textit{Proposition 1 (sufficient condition for tolerance-compliant reporting).}
Suppose that the participant is risk-neutral, cannot affect the authenticated reference or audit draw, and faces enforceable settlement. Reporting above the tolerance is weakly dominated if
\begin{equation}
\label{eq:general-compliance-condition}
\frac{\partial U_{i,j}^{t,r}(g)}{\partial g}\leq0, \qquad g\in(\varepsilon_j,\overline g_j].
\end{equation}

In the exact-reference limit $\nu_{i,z}^{t,j}=0$, let $\bar b_i^{t,j}\leq B_j$. A conservative parameter-level sufficient condition for Eq.~\eqref{eq:general-compliance-condition} is
\begin{equation}
\label{eq:exact-reference-compliance-condition}
\begin{aligned}
\delta^{\ell_j}p_j\pi_j(1+\eta\beta_P)
\geq{}&B_j+w_j(1-q_{E,j})\\
&+\delta^{\ell_j}q_{E,j}w_j(1-p_j).
\end{aligned}
\end{equation}

Under this condition, the expected utility is non\hyp{}increasing above $\varepsilon_j$, so no report beyond the tolerance can improve the participant's payoff. With noisy references, the exact functions $Q_j(g)$ and $\overline D_j(g)$ in Eq.~\eqref{eq:report-expected-utility-revised} must be used; Section~\ref{sec:behavioural-simulation} implements this case directly.
This result establishes conditional incentive compatibility within the stated evidence and enforcement assumptions. Blockchain does not establish the correctness of the off-chain reference. It preserves the report, committed parameters, due date, authenticated evidence pointer, audit outcome, and resulting consequence in one auditable chain.

This shared procedure applies to the Scene~1 records registered through Steps $3'$--$5'$ and to the event records in Scene~2. Its verified outputs are mapped either to governance\hyp{}oriented priority feedback in Step~$6'$ or to event-level settlement and bilateral credit. Thus, the two Chinese scenarios share data registration, evidence verification, credit updating, and rule execution while retaining different institutional outputs.

The proposed contribution-credit framework can also be adapted to the UK and US electricity markets while retaining the shared delayed-verification and reporting-penalty logic. The corresponding institutional mappings, workflow diagrams, and country-indexed formulations are provided in Appendices~A and~B.

\section{Controlled Heterogeneous-Agent Evaluation of Incentive Compatibility}
\label{sec:behavioural-simulation}
\noindent
This section evaluates a simplified instantiation of the Chinese mechanism, using Scene~1 as the representative setting. The simulation examines the behavioural-resource dimension of efficiency. Action compliance and chosen inflation measure reporting behaviour, effort and authenticated hard performance describe the coordination outcome, and transfer-neutral system surplus captures the outcome after effort, audit, physical-deviation, and report-distortion costs. Participant utility and reward--performance correlation are reported separately to expose trade-offs across objectives. The reported outcomes concern agent-level reporting and execution responses; they do not quantify the resource-allocation effects of Eq.~\eqref{eq:priority_update}. The experiment does not model the bilateral settlement process of Scene~2 or the institutional extensions to the UK and US markets, and it does not use confidential dispatch or settlement observations. Its results are controlled comparative statics under the stated calibration.

\subsection{Agent Decisions and Mechanism Ablation}
\label{subsec:agent-decisions-v21}
\noindent
Each agent represents a provincial, regional, or event\hyp{}level participant in the Chinese mechanism; regulators and headquarters\hyp{}level operators are not strategic agents. Agent $i$ has the fixed type $\theta_i=(b_i,h_i,q_i,c_i,s_i)$ within a replication, where $b_i$ is the private marginal benefit of inflation, $h_i$ the intrinsic marginal disutility of dishonest reporting, $q_i$ report-cost curvature, $c_i$ effort-cost curvature, and $s_i$ coordination productivity. The respective draws are
$\mathrm U(0.05,0.25)$, $\mathrm U(0.02,0.15)$,
$\mathrm U(0.25,0.45)$, $\mathrm U(0.60,1.00)$, and
$\mathrm U(0.85,1.05)$. Types are drawn once at the start of each replication;
only operating, measurement, commitment, and decision shocks vary over time.

In period $t$, agent $i$ chooses effort $e_i^t\in[0,1]$ and report inflation $g_i^t$. Under regime $r$, effort maximizes
\begin{equation}
\label{eq:sim-effort-utility-v21}
U_{i,r}^{t,e}(e)=m_{i,r}^t e-\frac{c_i}{2}e^2 -\mathbb{I}_{V,r}\pi_V \left[\bar y_i^t-s_i e-\tau_V\right]_+,
\end{equation}
where
\begin{equation}
\label{eq:sim-effort-margin-v21}
m_{i,r}^t=a_0s_i+\nu_{i,e}^t
+\mathbb{I}_{C,r}\omega_C\,Cr_i^t s_i.
\end{equation}
Here, $\mathbb{I}_{V,r}$ activates the execution-deviation penalty and $\mathbb{I}_{C,r}$ activates verified-credit feedback. 

The coefficients $a_0=0.61$ and $\omega_C=0.12$ are common across regimes; the credit term contributes to the marginal return only when $(\mathbb{I}_{C,r}=1$. This credit-to-effort term is a reduced-form behavioral assumption. The commitment target is $\bar y_i^t=\operatorname{clip}_{[0.55,0.90]}(0.72+\nu_{i,c}^t)$, the execution tolerance is $\tau_V=0.02$, and the execution-penalty intensity is $\pi_V=0.18$. The authenticated hard outcome and execution deviation are

\begin{equation}
\label{eq:sim-hard-outcome-v21}
y_i^t=\operatorname{clip}_{[0,1]} \left(s_i e_i^t+\omega_i^t\right),\qquad V_i^t=\left[\bar y_i^t-y_i^t-\tau_V\right]_+.
\end{equation}

Report inflation never enters $y_i^t$. Because the choice in Eq.~\eqref{eq:sim-effort-utility-v21} does not integrate over the future shock $\omega_i^t$, it is interpreted as a certainty-equivalent effort response.

The latent soft indicator is
\begin{equation}
x_i^t=\operatorname{clip}_{[0.05,0.95]}
\left[0.55+0.30(s_i-0.95)+\nu_{i,x}^t\right].
\label{eq:sim-latent-soft-v21}
\end{equation}
The report is $\widehat x_i^t=x_i^t+g_i^t$, where $0\leq g_i^t\leq\min\{0.35,1-x_i^t\}$. Let $\bar b_i^t=b_i-h_i+\nu_{i,g}^t$ and $\nu_{i,z}^t\sim\mathcal N(0,\sigma_m^2)$ with $\sigma_m=0.005$. For $a=g-\varepsilon$ and $k=a/\sigma_m$, the evidence law used when maximizing Eq.~\eqref{eq:report-expected-utility-revised} is
\begin{equation}
\label{eq:gaussian-deviation-moments-v21}
Q(g)=\Phi(k),\qquad
\overline D(g)=a\Phi(k)+\sigma_m\phi(k).
\end{equation}

Here, $\Phi(\cdot)$ and $\phi(\cdot)$ are the standard-normal distribution and density functions, respectively. The deterministic limits are $Q(g)=\mathbf 1\{a>0\}$ and $\overline D(g)=[a]_+$ when $\sigma_m=0$. The bounded one-dimensional maximizer evaluates global and Gaussian-transition candidates and refines the best bracket by golden-section search. Its action tolerance is $10^{-8}$. An independent spot check compared 256 decisions with 100,001-point grids; the maximum grid advantage in expected utility was zero, and the largest absolute action difference was $1.73\times10^{-6}$.

Verified credit is updated in two stages. Hard evidence produces
\begin{equation}
\label{eq:sim-hard-credit-v21}
\begin{aligned}
Cr_i^{t,H}={}&\operatorname{clip}_{[0,1]}\Big\{
Cr_i^t+(1-\rho)\beta_H(y_i^t-Cr_i^t)\\
&\qquad-\mathbb{I}_{V,r}\beta_VV_i^t
\Big\},
\end{aligned}
\end{equation}
while a soft report that becomes due at $t+\ell$ produces

\begin{equation}
\label{eq:sim-soft-credit-v21}
\begin{aligned}
Cr_i^{t+\ell,+}={}&\operatorname{clip}_{[0,1]}\Big\{
Cr_i^{t+\ell,-}
+(1-\rho)\beta_SA_i^t \\
&\quad\times(z_i^{t+\ell}-Cr_i^{t+\ell,-})
-\beta_PA_i^t\pi D_i^t \Big\}.
\end{aligned}
\end{equation}

Here, $Cr_i^{t,H}$ is credit after the period-$t$ hard-evidence
update, while $Cr_i^{t+\ell,-}$ and $Cr_i^{t+\ell,+}$ are credit
immediately before and after the due soft-report settlement. The
simulation uses $\beta_H=0.65$, $\beta_S=0.35$, $\beta_V=0.25$,
and $\beta_P=0.50$. Eqs.~\eqref{eq:sim-hard-credit-v21}
and~\eqref{eq:sim-soft-credit-v21} are the single-hard-indicator
and single-soft-indicator instantiation of
Eqs.~\eqref{eq:hard_credit_update} and
\eqref{eq:soft_credit_update}, with $y_i^t$ serving as the composite
authenticated hard-performance indicator.

Unaudited soft reports do not increase persistent credit. Reports
move through pending, due, and settled states and are settled exactly
once. At the due date, the simulator separately (i) forfeits the
escrow if an audited deviation is detected, (ii) subtracts
$A_i^t\pi D_i^t$ from the participant's payoff, and (iii) updates
credit by $-\beta_PA_i^t\pi D_i^t$.

For cohort-level reporting, the realized due-date terms are attributed
to their source reports, while the ledger retains the actual due and
settlement periods. Discounting affects the report decision and
discounted utility, not contract execution.

The existing action-compliance, chosen-inflation, realized\hyp{}penalty,
participant-payoff, discounted-utility, and reward--performance
results therefore already implement the definition above.
Transfer-neutral system surplus excludes penalties as transfers.
No simulation output is recalculated for this clarification.

\begin{table*}[!t]
\centering
\caption{Nested ablation regimes used in the controlled simulation.}
\label{tab:ablation-v21}
\scriptsize
\setlength{\tabcolsep}{4pt}
\renewcommand{\arraystretch}{1.08}
\begin{tabularx}{\textwidth}{p{0.16\textwidth}XXX}
\toprule
Regime & Record treatment & Execution and hard-credit treatment
& Soft-report treatment \\
\midrule
Baseline & Fragmented or ex-post records
& No execution penalty or verified-credit feedback
& No audit, escrow, or report-linked penalty \\
Blockchain only & Timestamped reports and evidence references
& Same payoff switches as Baseline
& Traceability alone has no direct utility term \\
Deviation penalty only & Traceable hard evidence
& Execution penalty and verified hard-credit feedback
& No audit, escrow, or report-linked penalty \\
Proposed verification penalty & Traceable hard and soft records
& Execution penalty and verified hard-credit feedback
& Delayed audit, persistent escrow, report penalty, and soft-credit update \\
\bottomrule
\end{tabularx}
\end{table*}

Table~\ref{tab:ablation-v21} separates record integrity from payoff-changing rules. Baseline and Blockchain only intentionally have identical behavioural switches: blockchain traceability without a consequence should not change the agent's choice in this model. Deviation penalty only activates execution accountability, whereas Proposed additionally activates delayed soft-report verification, escrow, and report-linked credit consequences.

\subsection{Simulation Setup and Evaluation Metrics}
\noindent
The experiment contains 100 agents and 72 action periods. The first 12 periods are discarded as a prespecified burn-in. Because the implemented credit recursion contains evidence weights and regime-specific feedback switches, its effective persistence is not simply $\rho^{12}$. A robustness check using the exported period-level paths and burn-in cutoffs of 18 and 24 periods changes the reported mean effort, hard performance, discounted utility, and system surplus by less than $3\times10^{-4}$. Each of 500 independent replications uses the same random bundle across regimes, enabling paired comparisons under common random numbers. Mean and rate outcomes are averaged over the 100 agents and 60 post-burn-in periods within each replication and then across replications. Reward--performance correlation is instead calculated from the pooled post-burn-in agent-period observations within each replication before being averaged across replications. The root seed is 20260716.

At the start of each period, reports due from earlier periods are settled before the current action is generated. The simulator then draws the current exogenous states and shocks, obtains the effort and reporting choices, realizes hard performance and current rewards, and registers the current soft report. When $\ell=0$, the newly registered report is settled in the same period. The hard-evidence credit update is then applied. After period 71, no new actions are generated, but the settlement tail continues until every pending report and escrow balance has been settled.

The fixed reward weights are 0.61 for hard performance, 0.14 for the soft report, and 0.25 for credit in every regime. The Proposed setting uses $Cr_i^0=0.50$, $\rho=0.60$, $p=0.70$, $\pi=0.30$, $\ell=1$, $\delta=0.90$, $\varepsilon=0.01$, a 50\% escrow share, and credit shadow value $\eta=0.20$. The standard deviations of the effort, report, latent-state, hard-performance, reference-measurement, and commitment shocks are 0.02, 0.015, 0.08, 0.04, 0.005, and 0.03, respectively. In the system-surplus calculation, the hard-performance value is 1.00, the cost per audit is 0.01, the physical-deviation cost is 0.40, and the report-distortion cost is 0.20. Sensitivity tests vary one of $p$, $\ell$, $\pi$, or the escrow share while holding the remaining parameters fixed.

The baseline values constitute an institutionally plausible calibration on the normalized $[0,1]$ scale rather than estimates obtained from confidential participant-level records. The reward weights of $0.61$, $0.14$, and $0.25$ assign the largest current-period weight to authenticated hard performance, limit the provisional influence of a soft report, and retain a persistent credit channel. The initial credit $Cr_i^0=0.50$ represents a neutral prior, while the commitment range $[0.55,0.90]$ and the productivity range $s_i\in[0.85,1.05]$ introduce moderate heterogeneity without causing frequent mechanical clipping at zero or one. The uncertainty magnitudes also preserve the intended evidence hierarchy: the reference-measurement standard deviation of $0.005$ is below the reporting tolerance $\varepsilon=0.01$ and is substantially smaller than the hard-performance and latent-state shocks of $0.04$ and $0.08$. The combination $p=0.70$, $\ell=1$, $\pi=0.30$, and a $50\%$ escrow share represents an enforceable but imperfect verification environment in which auditing is neither certain nor immediate. Because comparable participant-level audit and misreporting records are not publicly available, these values should not be interpreted as market estimates. The sensitivity ranges $p\in[0.30,0.90]$ and $\ell\in\{0,1,2,3\}$, together with the additional penalty-intensity and escrow-share tests, are therefore used to assess whether the conclusions depend on a single calibration.

Behaviour and evidence are measured separately. \emph{Action compliance} is $\Pr(g_i^t\leq\varepsilon)$, and \emph{chosen inflation} is $\mathbb E[g_i^t]$; both describe the agent's decision. \emph{Reference consistency} is $\Pr(D_i^t=0)$ after measurement noise, and \emph{realized deviation} is $\mathbb E[D_i^t]$; both describe the verification outcome. Reference consistency is not, by itself, labelled truthful reporting.

Nominal participant payoff attributes delayed settlements to the source report without discounting. Discounted participant utility multiplies delayed escrow release, report penalty, and the associated credit consequence by $\delta^\ell$. Transfer-neutral system surplus is an undiscounted resource measure: hard-performance value minus effort cost, audit cost, physical-deviation cost, and report-distortion cost. Incentive transfers, penalties, releases, and forfeitures are excluded. Consequently, any system-surplus comparison is conditional on the stated resource-cost specification and is not interchangeable with participant utility. Reward--performance correlation is treated as secondary because hard performance already enters the reward formula.

\subsection{Results and Incentive-Compatibility Implications}
\label{subsec:results-v21}

\begin{table*}[!t]
\centering
\caption{Controlled heterogeneous-agent simulation results. Values are means
across 500 replications; exact values and replication-level 95\% confidence
intervals are retained in the accompanying reproducibility package.}
\label{tab:main-results-v21}
\scriptsize
\setlength{\tabcolsep}{5pt}
\renewcommand{\arraystretch}{1.08}
\resizebox{\textwidth}{!}{%
\begin{tabular}{lcccc}
\toprule
Outcome & Baseline & Blockchain only & Deviation penalty only
& Proposed verification penalty \\
\midrule
Action compliance & $<0.001$ & $<0.001$ & $<0.001$ & 0.998 \\
Chosen inflation & 0.334 & 0.334 & 0.334 & 0.001 \\
Reference consistency & $<0.001$ & $<0.001$ & $<0.001$ & 0.975 \\
Realized deviation & 0.324 & 0.324 & 0.324 & 0.001 \\
Coordination effort & 0.740 & 0.740 & 0.865 & 0.858 \\
Hard performance & 0.704 & 0.704 & 0.819 & 0.813 \\
Discounted participant utility & 0.466 & 0.466 & 0.616 & 0.533 \\
Transfer-neutral system surplus & 0.402 & 0.402 & 0.457 & 0.513 \\
Reward--performance correlation & 0.991 & 0.991 & 0.987 & 0.983 \\
\bottomrule
\end{tabular}%
}
\end{table*}

\noindent
Table~\ref{tab:main-results-v21} reports the principal controlled comparisons. Baseline and Blockchain only are identical for every replication and reported outcome. This is the intended falsification result: record traceability alone does not change behaviour when it has no payoff consequence. In contrast, the Proposed regime produces action compliance of 0.998462 (95\% CI 0.998312--0.998612) and mean chosen inflation of 0.000606. After reference-measurement noise, consistency is 0.974845 and mean realized deviation is 0.000552. Relative to Baseline, the paired changes are $+0.998455$ for action compliance (95\% CI 0.998305--0.998605) and $-0.333349$ for chosen inflation (95\% CI $-0.333642$--$-0.333055$). Under the frozen payoff, evidence, and enforcement assumptions, delayed verification therefore changes the simulated reporting decision; it does not establish honesty in a field population.

The Proposed regime also raises mean effort from 0.739\allowbreak661 to 0.858202 and hard performance from 0.704171 to 0.812808 relative to Baseline. These differences cannot be attributed to blockchain recording, because Blockchain only is identical to Baseline. Nor do they show that report verification maximises coordination: Deviation penalty only produces effort of 0.865037 and hard performance of 0.819012. Relative to that regime, the paired Proposed effects are $-0.006835$ for effort (95\% CI $-0.006892$--$-0.006779$) and $-0.006204$ for hard performance (95\% CI $-0.006256$--$-0.006152$).

Discounted participant utility is 0.532897 under the Proposed regime, higher than
Baseline (0.466469) but lower than Deviation penalty only (0.616090). Proposed
has the largest transfer-neutral system surplus, 0.513371, compared with
0.401781 under Baseline and 0.457385 under Deviation penalty only. This result
is conditional on treating reporting distortion as a resource cost; it is not
interchangeable with participant utility. Reward--performance correlation is
also lower under Proposed (0.983430) than under Baseline (0.990914) or
Deviation penalty only (0.986516). The simulation therefore supports neither
a correlation-improvement claim nor universal dominance of the Proposed
mechanism.

\begin{table}[!t]
\centering
\caption{Sensitivity to audit probability and verification delay.}
\label{tab:sensitivity-v21}
\scriptsize
\setlength{\tabcolsep}{3pt}
\renewcommand{\arraystretch}{1.06}
\resizebox{\columnwidth}{!}{%
\begin{tabular}{lcccc}
\toprule
Setting & Action compliance & Chosen inflation & Discounted utility
& System surplus \\
\midrule
$p=0.3$ & 0.452 & 0.177 & 0.548 & 0.483 \\
$p=0.5$ & 0.872 & 0.043 & 0.537 & 0.507 \\
$p=0.7$ & 0.998 & 0.001 & 0.533 & 0.513 \\
$p=0.9$ & 1.000 & $<0.001$ & 0.527 & 0.511 \\
$\ell=0$ & 1.000 & $<0.001$ & 0.544 & 0.514 \\
$\ell=1$ & 0.998 & 0.001 & 0.533 & 0.513 \\
$\ell=2$ & 0.990 & 0.004 & 0.529 & 0.513 \\
$\ell=3$ & 0.967 & 0.011 & 0.525 & 0.511 \\
\bottomrule
\end{tabular}%
}
\end{table}

Table~\ref{tab:sensitivity-v21} reports the main-text sensitivity to audit probability and verification delay; penalty-intensity and escrow-share results are retained in the reproducibility package. Increasing $p$ from 0.3 to 0.9 raises action compliance from 0.451811 to 1.0 and reduces chosen inflation from 0.177144 to 0.000012. Increasing $\ell$ from 0 to 3 lowers action compliance from 0.999935 to 0.966953 and raises chosen inflation from 0.000072 to 0.011220. Stronger auditing does not monotonically increase every outcome: discounted utility falls as $p$ increases, and system surplus is slightly lower at $p=0.9$ than at $p=0.7$ because verification is costly. The sensitivity analysis therefore identifies a design trade-off rather than implying that maximal auditing is always optimal.

\begin{figure*}[!t]
\centering
\begin{minipage}[t]{0.235\textwidth}
\centering
\includegraphics[width=\linewidth]{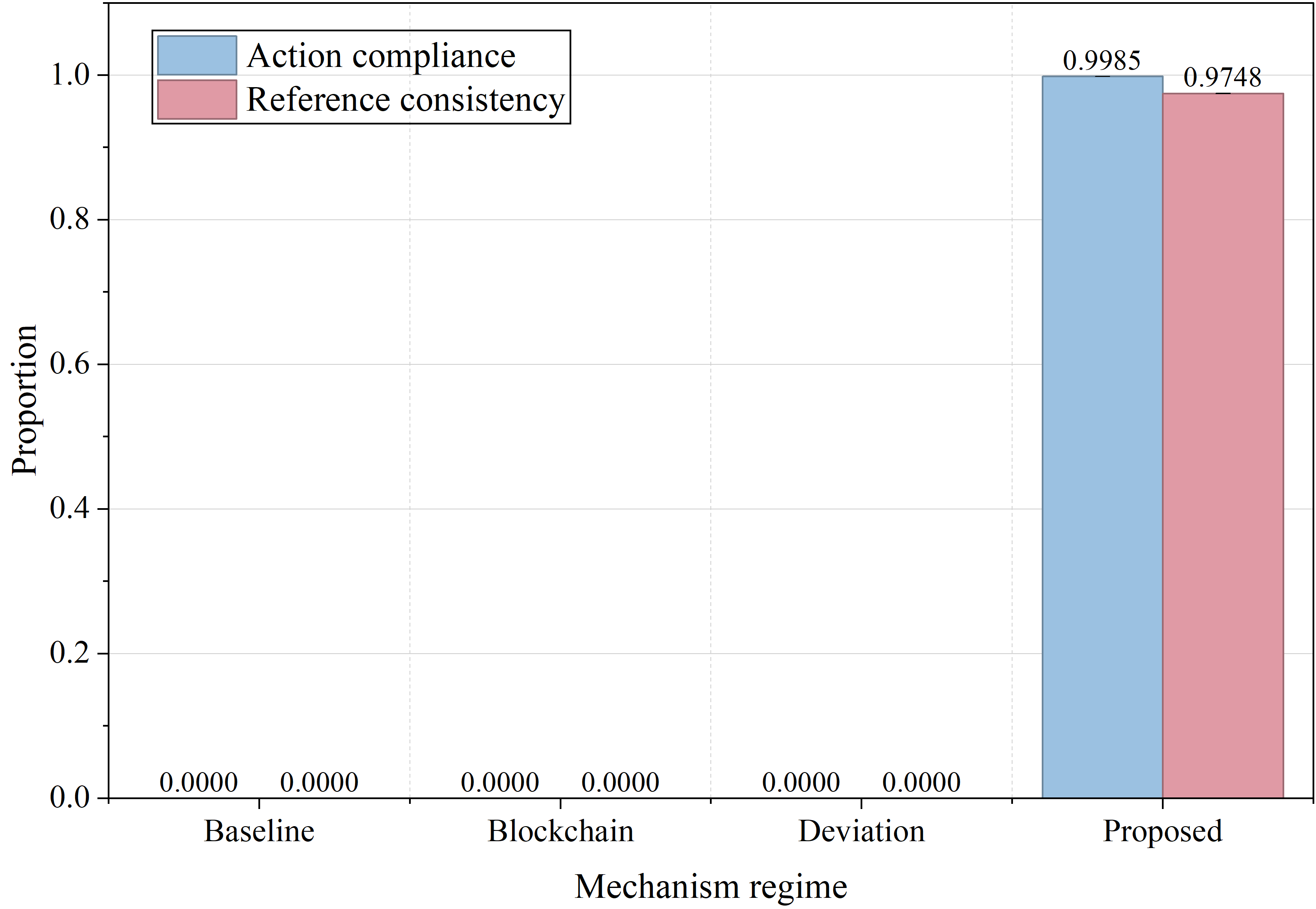}
\par\smallskip
{\scriptsize (a) Action compliance and reference consistency
(proportions).\par}
\end{minipage}\hfill
\begin{minipage}[t]{0.235\textwidth}
\centering
\includegraphics[width=\linewidth]{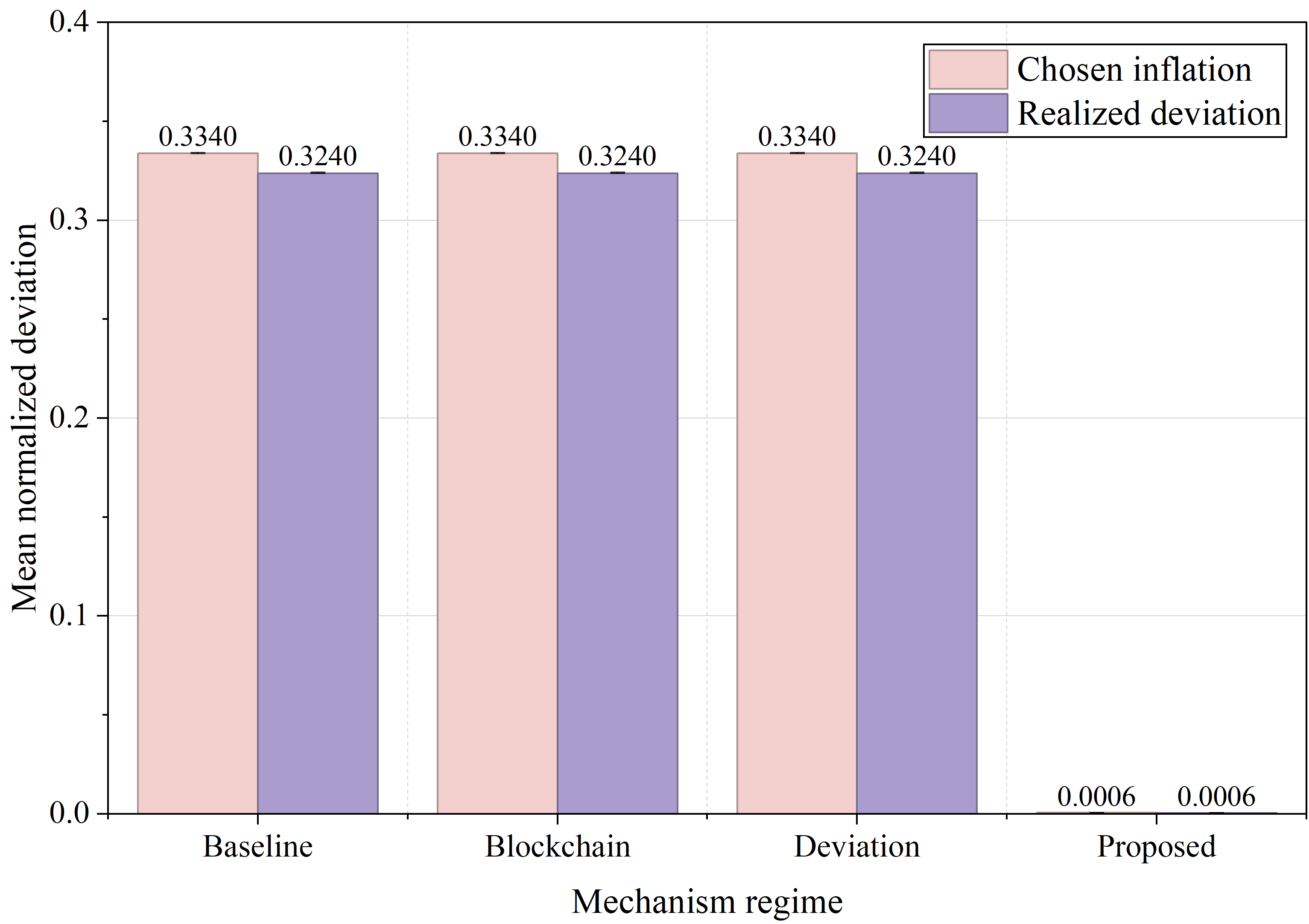}
\par\smallskip
{\scriptsize (b) Chosen inflation and realized deviation
(mean normalized values).\par}
\end{minipage}\hfill
\begin{minipage}[t]{0.235\textwidth}
\centering
\includegraphics[width=\linewidth]{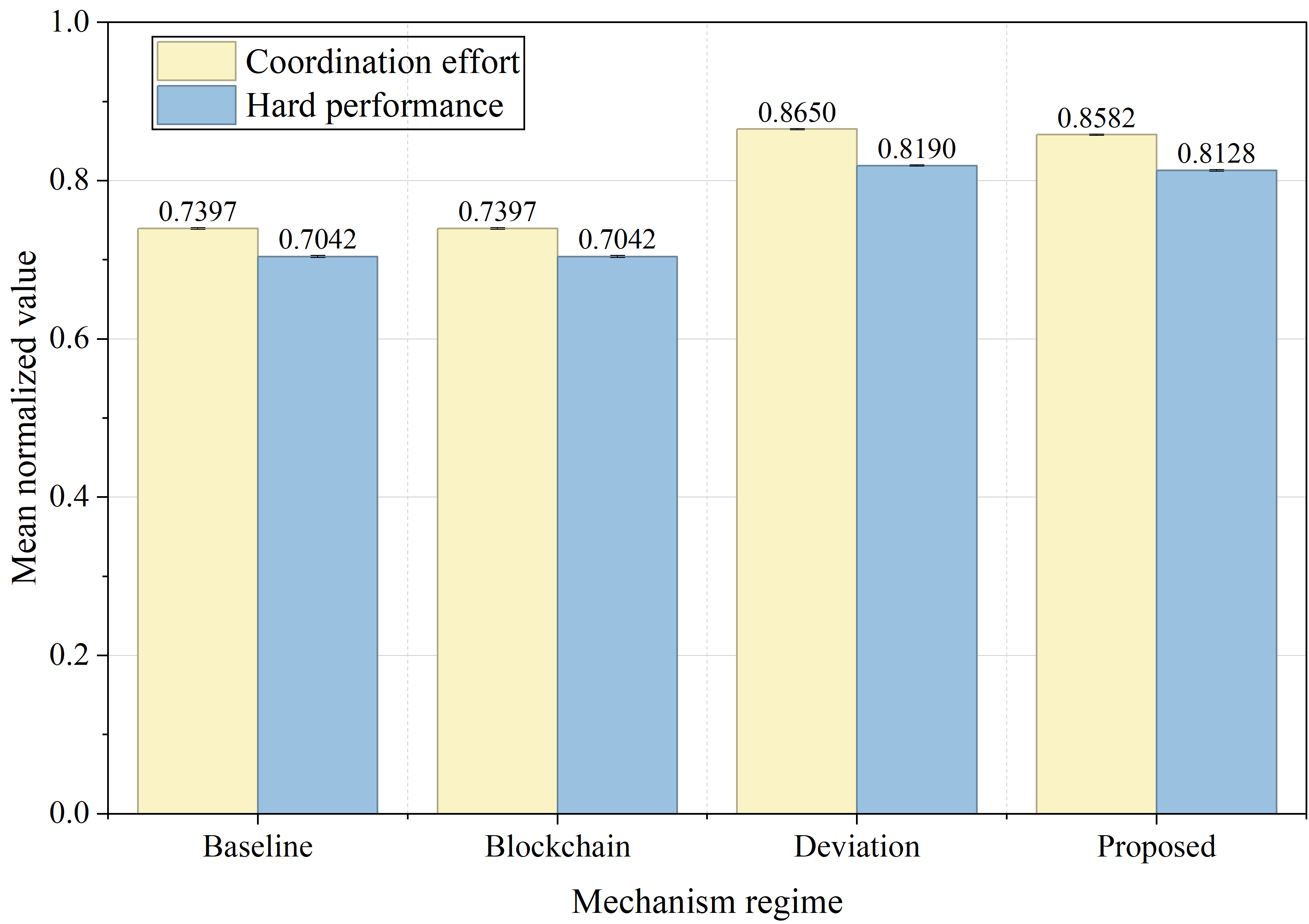}
\par\smallskip
{\scriptsize (c) Coordination effort and authenticated hard
performance (mean normalized values).\par}
\end{minipage}\hfill
\begin{minipage}[t]{0.235\textwidth}
\centering
\includegraphics[width=\linewidth]{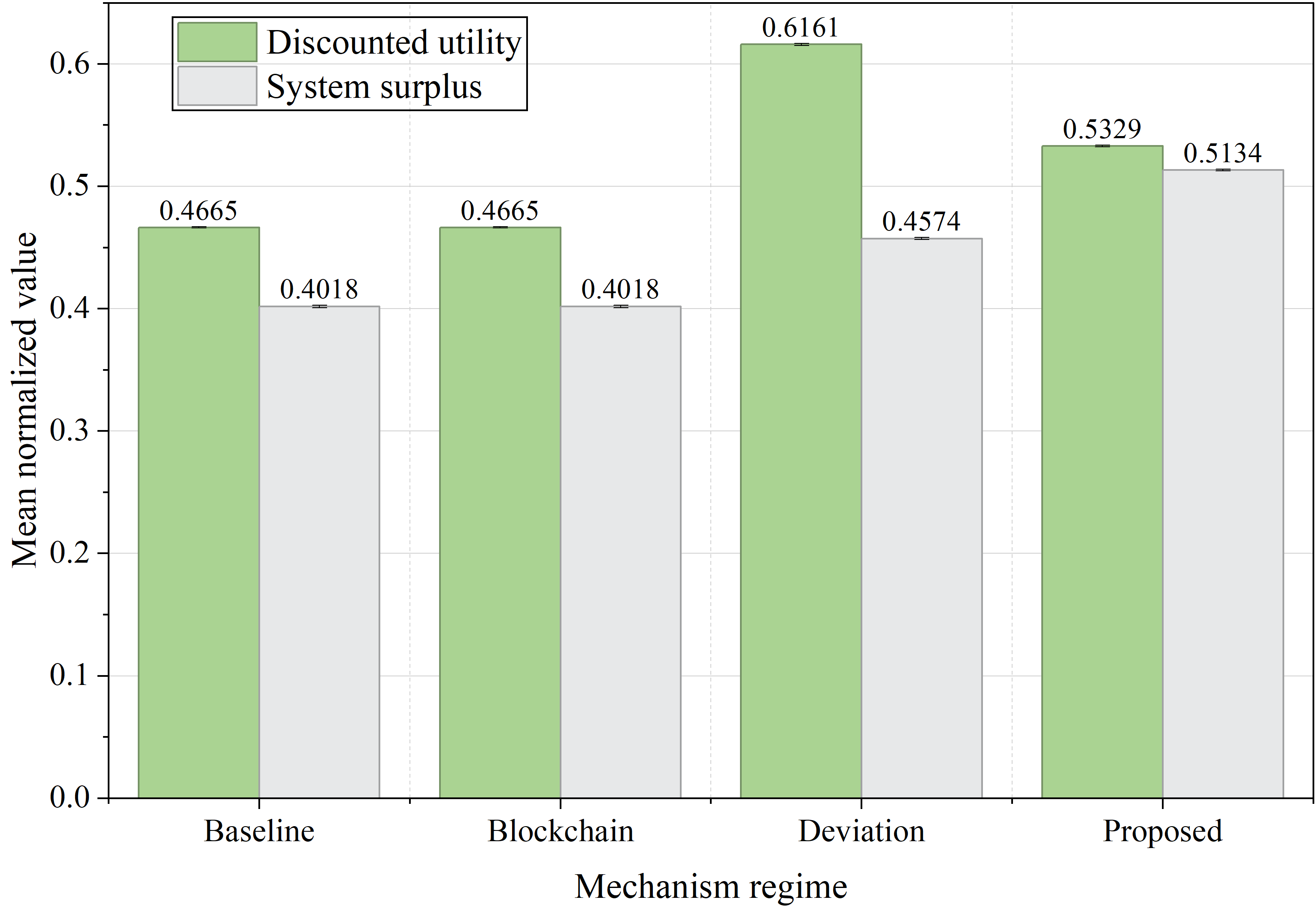}
\par\smallskip
{\scriptsize (d) Discounted participant utility and transfer-neutral
system surplus (mean normalized values).\par}
\end{minipage}

\caption{Controlled heterogeneous-agent simulation across mechanism
regimes (x-axis). Error bars indicate replication-level 95\%
confidence intervals. Panel (d) uses a focused vertical scale;
exact values are reported in Table~\ref{tab:main-results-v21}.}
\label{fig:behavior_simulation}
\end{figure*}

Fig.~\ref{fig:behavior_simulation} visualises the separation among reporting choices, ex-post evidence, physical performance, participant utility, and system surplus. Under the stated normalized assumptions, blockchain traceability alone changes neither reporting choices nor coordination outcomes. Linking later evidence to escrow, report-linked penalties, and persistent credit sharply reduces report inflation and produces the highest transfer-neutral system surplus because the reduction in distortion and deviation costs exceeds the additional audit cost. The Deviation penalty only regime still produces slightly higher effort, hard performance, participant utility, and reward--performance correlation. Accordingly, the supported efficiency result is dimension-specific: the Proposed mechanism improves reporting compliance and distortion-adjusted resource efficiency under the specified cost structure, while preserving visible trade-offs across participant and operational outcomes. These results remain controlled comparative statics of the mechanism and do not constitute actor-level empirical estimates.

\section{Security Analysis, China Data Anchoring, and Execution Benchmarks}

\subsection{Security Analysis}
\noindent
The analysis follows the three requirements used to construct the mechanism: shared coordination, verifiable performance recognition, and rule-based incentive execution. These requirements are treated as security gates rather than as substitutable performance dimensions. A correct ledger state cannot compensate for unreliable source evidence, and reliable evidence cannot compensate for an incorrect or replayed contract execution.

The threat model includes a participant that inflates a soft report, an insider or external party that alters or replays an off-chain record after registration, an unauthorised party that requests protected data, and a transaction that is invoked with incomplete or stale evidence. The analysis assumes collision-resistant hashing, unforgeable digital signatures, authenticated consortium identities, deterministic contract logic, and ledger consensus within the platform's stated fault bound. It does not assume that a blockchain can establish the truth of an off-chain measurement. Correlated source errors or collusion among enough authorised evidence providers remain outside the guarantee and must trigger institutional review.

Let $r_i^{t,j}$ be the off-chain source record for actor $i$, reporting period $t$, and indicator $j$. Its metadata $m_i^{t,j}$ contains at least the actor, period, indicator, timestamp, and record version. The on-chain commitment is $H_i^{t,j}=h(r_i^{t,j}\Vert m_i^{t,j})$, accompanied by the source signature $\sigma_i^{t,j}$. As in Section~3, the country index is omitted for the Chinese mechanism.

\subsubsection{Shared coordination security: ledger integrity and permissioned visibility}
\noindent
The consortium ledger shares commitments, verification states, and rule outputs; detailed metering, dispatch, settlement, and qualification files remain in their source systems. For a tested access event $(u,i,j)$ in period $t$, let $p_{u,i}^{t,j}\in\{0,1\}$ be the policy-prescribed decision for requester $u$, and let $d_{u,i}^{t,j}$ be the decision enforced by the access-control contract. Record integrity, source authenticity, and policy enforcement are tested jointly by
\begin{subequations}\label{eq:ledger-security}
\begin{align}
\mathcal{L}_{u,i}^{t,j}
={}&\mathbf{1}\!\left[
h(r_i^{t,j}\Vert m_i^{t,j})=H_i^{t,j}
\right] \notag\\
&\times\mathbf{1}\!\left[
\operatorname{Verify}_{pk_i}
(\sigma_i^{t,j},H_i^{t,j})=1
\right] \notag\\
&\times\mathbf{1}\!\left[
d_{u,i}^{t,j}=p_{u,i}^{t,j}
\right],
\label{eq:record-security-gate}\\
\mathcal{S}^{t}
={}&\frac{1}{|\mathcal{E}^{t}|}
\sum_{(u,i,j)\in\mathcal{E}^{t}}
\mathcal{L}_{u,i}^{t,j},
\label{eq:ledger-consistency-score}
\end{align}
\end{subequations}

where $\mathcal{E}^{t}$ contains the access and verification events tested in period $t$, including both permitted and denied requests. This denominator avoids treating a legitimate denial as a consistency failure. The period passes the shared-record gate only when $\mathcal{S}^{t}=1$; an average below one is a diagnostic rate, not a level that can be offset by another security dimension. Binding the identity, period, indicator, and version into the commitment also makes substitution and replay detectable. Subject to the assumed consensus fault bound, authorised peers then observe the same committed state and verification status.

\subsubsection{Verifiable performance security: evidence reliability and oracle trust}
\noindent
The oracle boundary connects off-chain electricity systems to the contract. Consensus can preserve the submitted evidence, but it cannot independently prove that a meter reading, dispatch instruction, qualification result, or settlement quantity was correct at source \cite{hassan2023oracle}. For indicator $j$, let $\mathcal{O}_j$ contain the approved evidence channels and let $z_{i,\ell}^{t+\ell_j,j}$ be the normalized value returned by channel $\ell$ when the report from period $t$ becomes verifiable. The channels may include metering systems, dispatch logs, settlement records, and regulatory audit files.

Let $q_{j,\ell}^{t+\ell_j}\in[0,1]$ be the current reliability rating of channel $\ell$, and let $\omega_{j,\ell}\geq0$, with $\sum_{\ell\in\mathcal{O}_j}\omega_{j,\ell}=1$, be a weight fixed before evidence is observed. Define the effective weight $\vartheta_{j,\ell}^{t+\ell_j}=\omega_{j,\ell}q_{j,\ell}^{t+\ell_j}$. Let $\bar z_i^{t+\ell_j,j}$ be the output of an aggregation rule registered for indicator $j$ before reports are submitted; if policy designates a single authoritative source, it is simply that source value. The binary channel-consistency test is $\mathcal{C}_{i,\ell}^{t,j}=\mathbf{1}[|z_{i,\ell}^{t+\ell_j,j}-\bar z_i^{t+\ell_j,j}|\leq\varepsilon_j]$. Evidence reliability and its acceptance gate are
\begin{subequations}\label{eq:evidence-security}
\begin{align}
ER_i^{t,j}
={}&\sum_{\ell\in\mathcal{O}_j}
\vartheta_{j,\ell}^{t+\ell_j}
\mathcal{C}_{i,\ell}^{t,j},
\label{eq:evidence-reliability}\\
\mathcal{V}_i^{t,j}
={}&\mathbf{1}\!\left[ER_i^{t,j}\geq\tau_j\right]
\mathbf{1}\!\left[
\sum_{\ell\in\mathcal{O}_j}
\mathcal{C}_{i,\ell}^{t,j}
\geq m_j
\right].
\label{eq:evidence-acceptance-gate}
\end{align}
\end{subequations}

Here, $\tau_j$ is the minimum accepted reliability, and $m_j$ is the required number of mutually consistent channels. The aggregation rule, weights, threshold, and channel membership must be registered before verification. If $\mathcal{V}_i^{t,j}=1$, the authenticated reference in Section~\ref{subsec:shared-delayed-verification} is set to $z_i^{t+\ell_j,j}=\bar z_i^{t+\ell_j,j}$. Otherwise, the report remains pending or is routed to manual audit and cannot increase persistent credit or trigger final settlement. This rule avoids the circular test of comparing a source with an indicator already labelled as verified. It cannot protect against correlated errors or collusion sufficient to control the accepted evidence set.

\subsubsection{Automatically executable incentive security: deterministic and bounded execution}
\noindent
An execution item $e$ may be a Scene~1 contribution, credit, or priority update, or a Scene~2 event settlement and bilateral credit update. Let $\mathcal{K}_e$ be the record-access events required by item $e$, and let $\mathcal{Q}_e$ be its required evidence items. The pre-execution gate and the execution-correctness gate are
\begin{subequations}\label{eq:execution-security}
\begin{align}
\mathcal{G}_e
={}&\prod_{(u,i,t,j)\in\mathcal{K}_e}
\mathcal{L}_{u,i}^{t,j}
\prod_{(i,t,j)\in\mathcal{Q}_e}
\mathcal{V}_i^{t,j},
\label{eq:pre-execution-gate}\\
\mathcal{X}_e
={}&\mathbf{1}\!\left[\mathbf{y}_e=\mathbf{y}_e^{*}\right]
\mathbf{1}\!\left[L_e^{\mathrm{exec}}\leq\bar L\right]
\mathbf{1}\!\left[n_e=n_e^{-}+1\right].
\label{eq:execution-correctness-gate}
\end{align}
\end{subequations}

Here, $\mathbf{y}_e$ is the vector of contract outputs and $\mathbf{y}_e^{*}$ is independently recomputed from the same verified state; $L_e^{\mathrm{exec}}$ is execution latency, $\bar L$ is its operational bound, and $n_e$ is a monotonically increasing state version or nonce. The nonce test prevents duplicate or stale execution. Exact equality is appropriate when the contract uses deterministic fixed-point arithmetic; an implementation using independently rounded real-valued calculations must register a componentwise tolerance in advance.

The item-level security decision is deliberately non-compensatory:
\begin{equation}
\label{eq:noncompensatory-security-gate}
\mathcal{R}_e=\mathcal{G}_e\mathcal{X}_e.
\end{equation}

Thus, $\mathcal{R}_e=1$ only when every required record passes the integrity and access-control checks, every required evidence item passes the oracle gate, and the resulting state transition is correct, timely, and non-replayed. If any component is zero, the contract withholds the final incentive consequence and records the exception for authorised review. The quantities $\mathcal{S}^t$ and $ER_i^{t,j}$ may be reported as diagnostic rates, but they are not combined into a weighted ``security score.'' This distinction prevents a severe oracle or execution failure from being hidden by a high average in another dimension. The same gates apply to the country-indexed extensions in Appendices~A and~B after replacing the institutional evidence sets and contract outputs.

\subsection{China Public-Data Anchoring and Workflow Analysis}
\label{subsec:china-workflow-analysis}
\noindent
The evidence in this paper is intentionally separated by function. Section~\ref{sec:behavioural-simulation} tests simulated reporting and effort choices under alternative payoff rules. The present subsection instead establishes the scale and institutional relevance of the two Chinese scenarios and compares how records move through the existing and proposed workflows. The blockchain benchmarks in the following subsection then test implementation overhead. Aggregate public statistics are not treated as actor-level contribution records or as causal evidence of the mechanism's effect.

China publishes authoritative aggregate statistics on renewable-energy operation and interregional electricity trading, but public actor-level data do not jointly identify submitted commitments, authenticated execution, delayed verification, credit consequences, and mutual-aid settlement. Table~\ref{tab:china-public-anchors} therefore uses official 2024 statistics only as scale anchors. The National Energy Administration reports 3.46~PWh of renewable generation, approximately 35\% of total generation, wind and photovoltaic utilization rates of 95.9\% and 96.8\%, and 1.40~PWh of cross-provincial and cross-regional market-based electricity trading.\footnote{National Energy Administration, ``2024 Renewable Energy Grid-connected Operation,'' Jan. 2025, \url{https://www.nea.gov.cn/20250221/e10f363cabe3458aaf78ba4558970054/c.html}.}\footnote{National Energy Administration, ``Ten Years of the New Round of Power-sector Reform,'' Mar. 2025, \url{https://www.nea.gov.cn/20250321/e28b932e3ffb41aa9ab84b8efffba11c/c.html}.}

\begin{table}[!t]
\centering
\caption{Official 2024 China Data Used as Scenario-scale Anchors}
\label{tab:china-public-anchors}
\scriptsize
\setlength{\tabcolsep}{2.5pt}
\renewcommand{\arraystretch}{1.08}
\begin{tabularx}{\columnwidth}{
  >{\raggedright\arraybackslash}p{0.30\columnwidth}
  >{\raggedright\arraybackslash}p{0.22\columnwidth}
  Y}
\toprule
Public indicator & Anchor & Role in this study \\
\midrule
Renewable generation
& 3.46 PWh; about 35\% of generation
& Establishes the scale of renewable accommodation. \\
Wind and PV utilization
& 95.9\%; 96.8\%
& Motivates fulfilment and deviation indicators. \\
Interregional market-based trading
& 1.40 PWh
& Establishes the relevance of cross-provincial coordination. \\
\bottomrule
\end{tabularx}
\end{table}

These anchors determine neither the normalized actor-level inputs nor the numerical outcomes of the behavioural simulation. Their role is limited to showing that renewable accommodation and interregional coordination are material system functions. The hybrid AHP--entropy procedure in Appendix~C is likewise an illustrative parameter-selection method, not a structural estimate from the aggregate statistics.

Because a common event-level dataset linking all required records is unavailable, this subsection does not report estimated baseline-to-mechanism percentage gains. Instead, Table~\ref{tab:china-workflow-comparison} compares the two Chinese workflows at the rule and evidence level. The existing workflow is a stylized fragmented-record reference consistent with Fig.~\ref{fig:china}(a), rather than a claim that every current Chinese process lacks digital verification. The proposed workflow corresponds to Fig.~\ref{fig:china}(b) and Eqs.~\eqref{eq:normalization}--\eqref{eq:noncompensatory-security-gate}.

\begin{table*}[!t]
\centering
\caption{Rule-level Comparison of the China Coordination Workflows}
\label{tab:china-workflow-comparison}
\scriptsize
\setlength{\tabcolsep}{4pt}
\renewcommand{\arraystretch}{1.10}
\begin{tabularx}{\textwidth}{p{0.16\textwidth}XXp{0.22\textwidth}}
\toprule
Function & Fragmented-record reference workflow & Proposed verified workflow & Observable validation criterion \\
\midrule
Record registration
& Coordination, dispatch, metering, and settlement records remain in separate institutional systems.
& Signed commitments and evidence pointers are time-stamped and linked by actor, period, indicator, and version.
& Hash and signature checks in $\mathcal L_{u,i}^{t,j}$. \\
Evidence treatment
& Reconciliation is performed after records are exchanged across organizational boundaries.
& Hard evidence is authenticated directly; soft reports remain pending until the registered reference becomes available.
& Evidence gate $\mathcal V_i^{t,j}$ and due-date state transition. \\
Strategic reports
& A favorable claim may affect an administrative calculation before its later evidence is linked to the original report.
& Provisional rewards are escrowed, and audited deviations trigger $\Pi_i^{\tau}$ and future-credit consequences.
& Report--reference deviation $D_i^{t,j}$ and exactly-once settlement. \\
Contribution and credit
& Performance recognition and historical reliability may be maintained in separate assessment processes.
& Finalized contribution and persistent credit are calculated separately from authenticated evidence.
& Recalculation of $C_i^t$ and the two-stage $Cr_i$ update. \\
Incentive execution
& Priority or settlement changes require additional institutional reconciliation.
& A rule output is released only when every required record, evidence item, and state transition passes its gate.
& Non-compensatory decision $\mathcal R_e=1$. \\
Exception handling
& Disputed or incomplete records are resolved through the applicable administrative or settlement process.
& Failed gates suspend execution and preserve an auditable exception for authorized review.
& Logged rejection reason, nonce, and review status. \\
\bottomrule
\end{tabularx}
\end{table*}

This comparison establishes what the mechanism changes and how each change can later be tested; it does not assign an empirical effect size. The behavioural experiment addresses the participant-choice channel, while the following blockchain benchmarks address computational execution. UK and US public-data context is retained in Appendix~D so that the main analysis remains centered on China.

At the workflow level, coordination-process efficiency refers to the integration of evidence-linked state transitions. Table~7 shows that signed reports, authenticated evidence, delayed settlement, credit consequences, and final incentive outputs are connected by actor, period, indicator, and verification state. This structure reduces unlinked reconciliation points and makes each transition independently testable. Because a common event-level dataset on processing time and administrative cost is unavailable, the workflow evidence supports process integration and accountability without assigning a numerical time-saving or cost-reduction effect.

\subsection{Blockchain Execution and Permissioned-network Benchmark}

\begin{figure*}[!tbp]
\centering
\begin{minipage}[t]{0.235\textwidth}
\centering
\includegraphics[width=\linewidth]{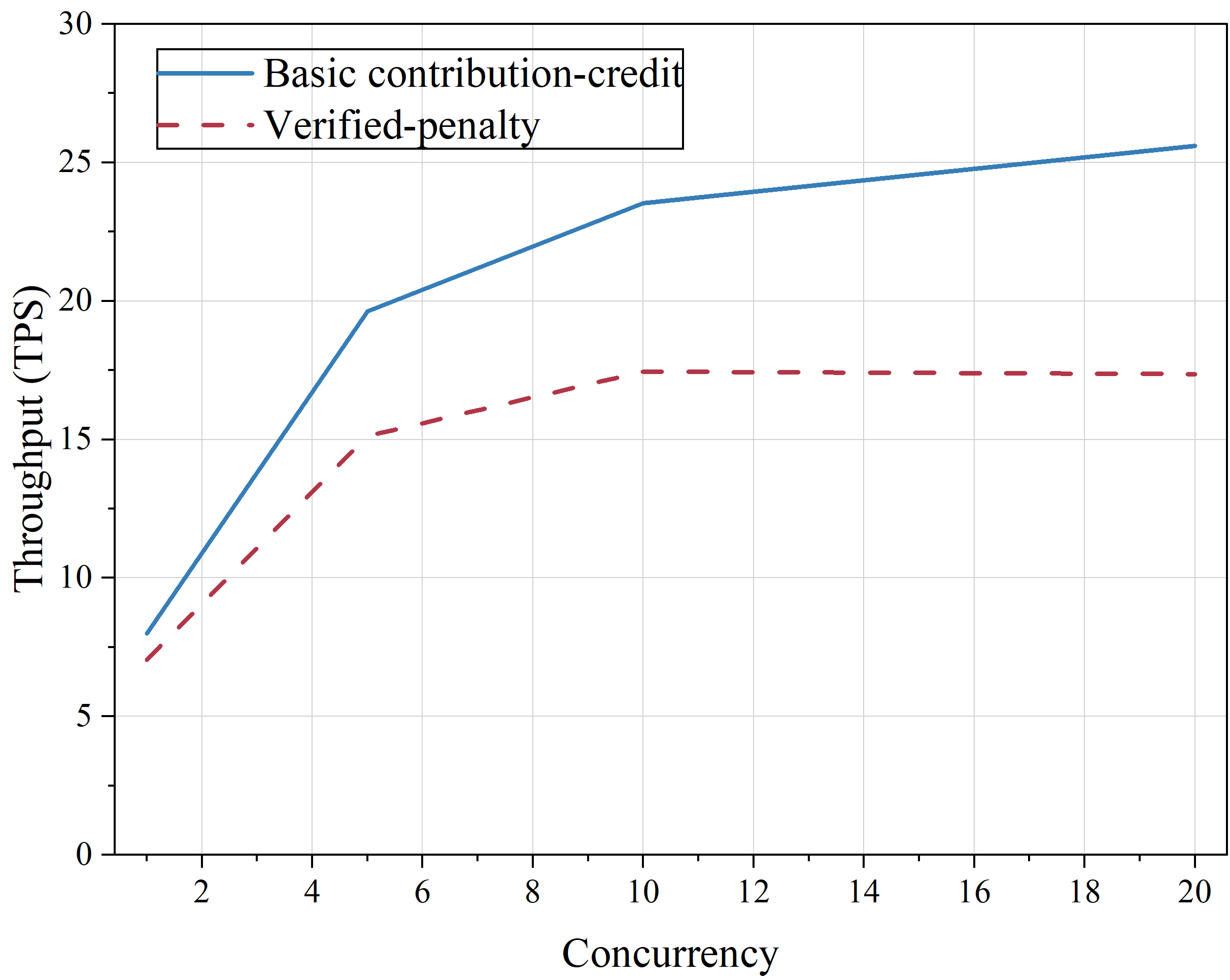}
\par\smallskip
{\scriptsize (a) Ganache throughput (TPS): basic versus
verified-penalty contract.\par}
\end{minipage}\hfill
\begin{minipage}[t]{0.235\textwidth}
\centering
\includegraphics[width=\linewidth]{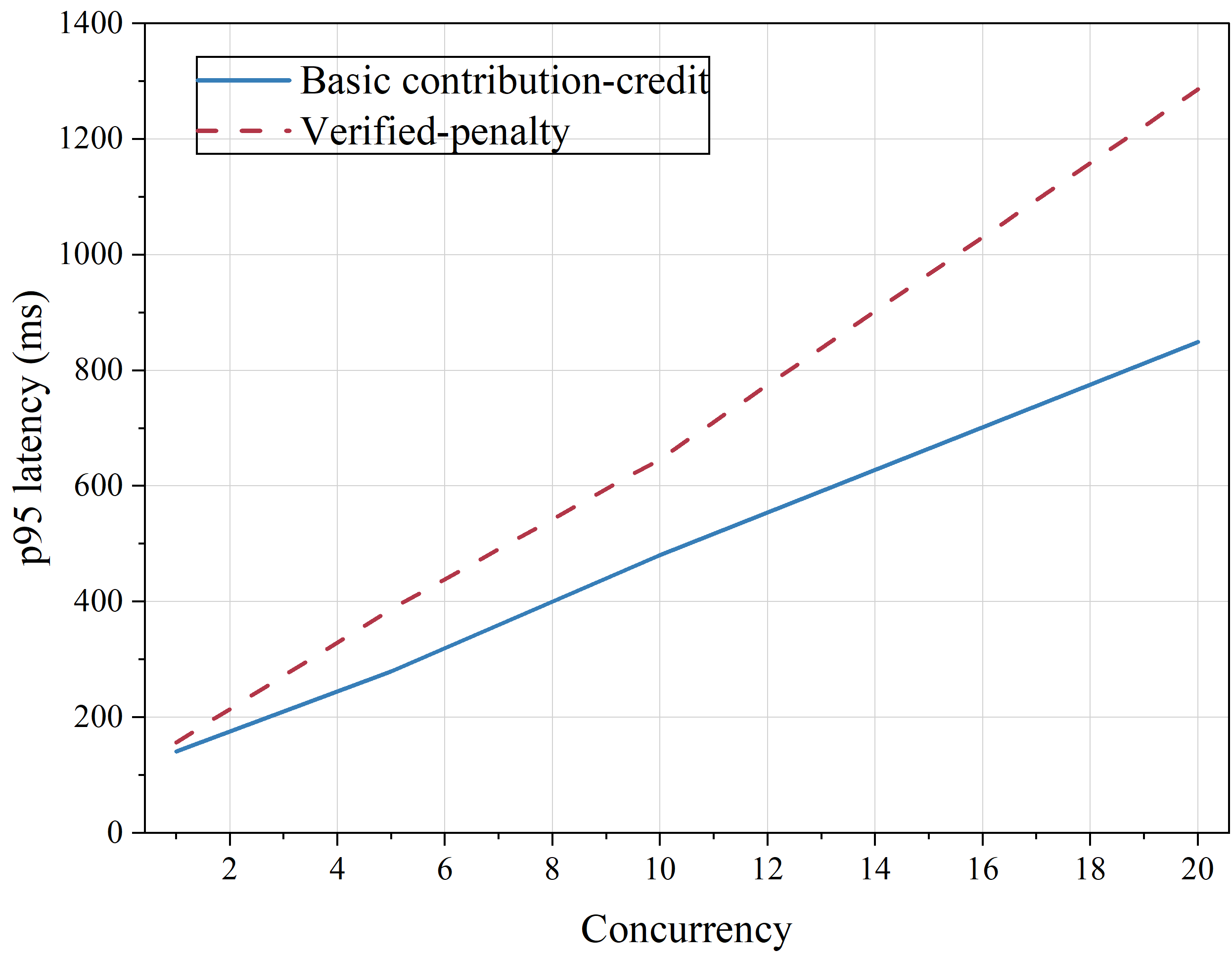}
\par\smallskip
{\scriptsize (b) Ganache p95 latency: basic versus
verified-penalty contract.\par}
\end{minipage}\hfill
\begin{minipage}[t]{0.235\textwidth}
\centering
\includegraphics[width=\linewidth]{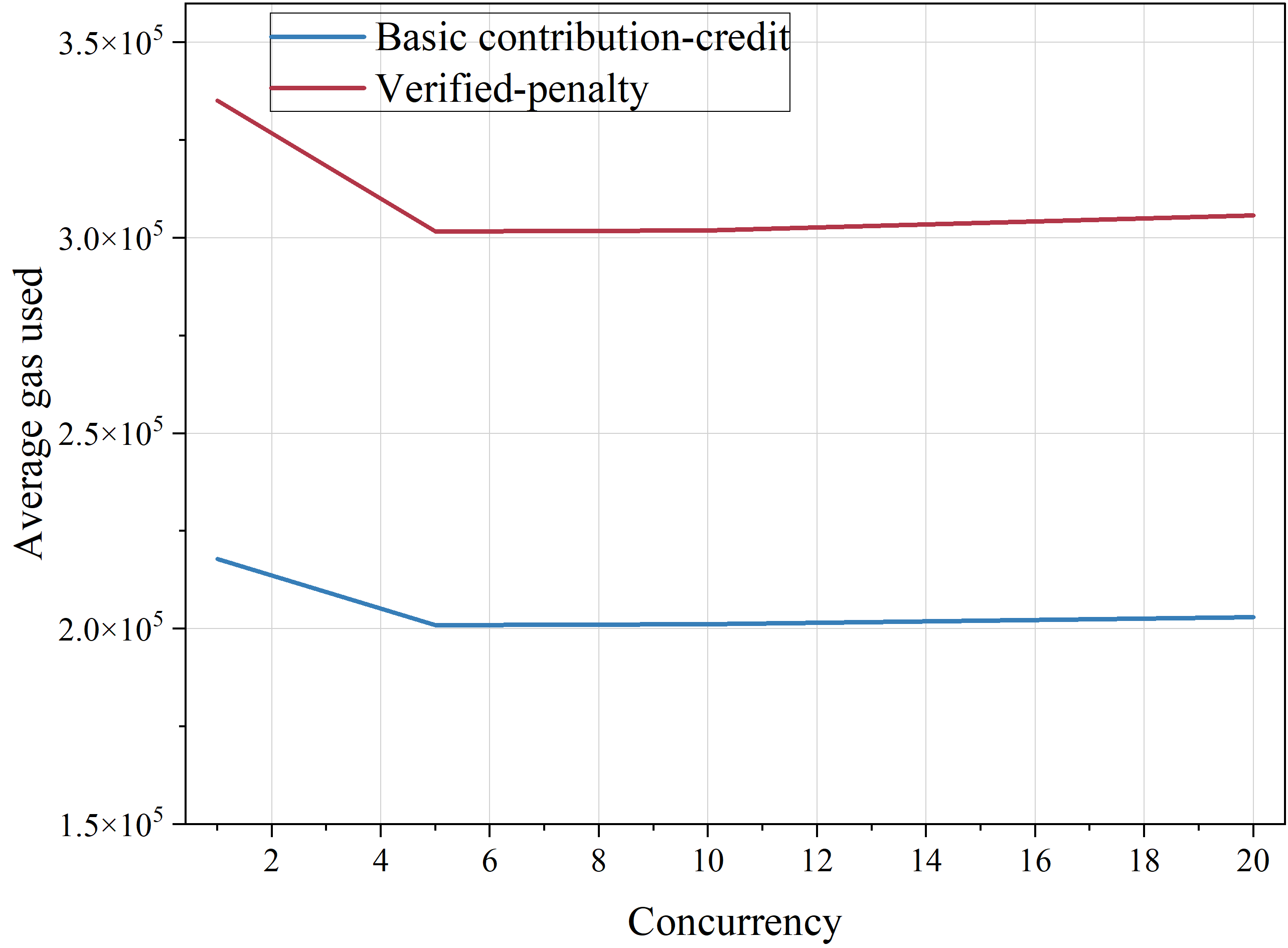}
\par\smallskip
{\scriptsize (c) Average gas consumption per transaction
for the two contracts.\par}
\end{minipage}\hfill
\begin{minipage}[t]{0.235\textwidth}
\centering
\includegraphics[width=\linewidth]{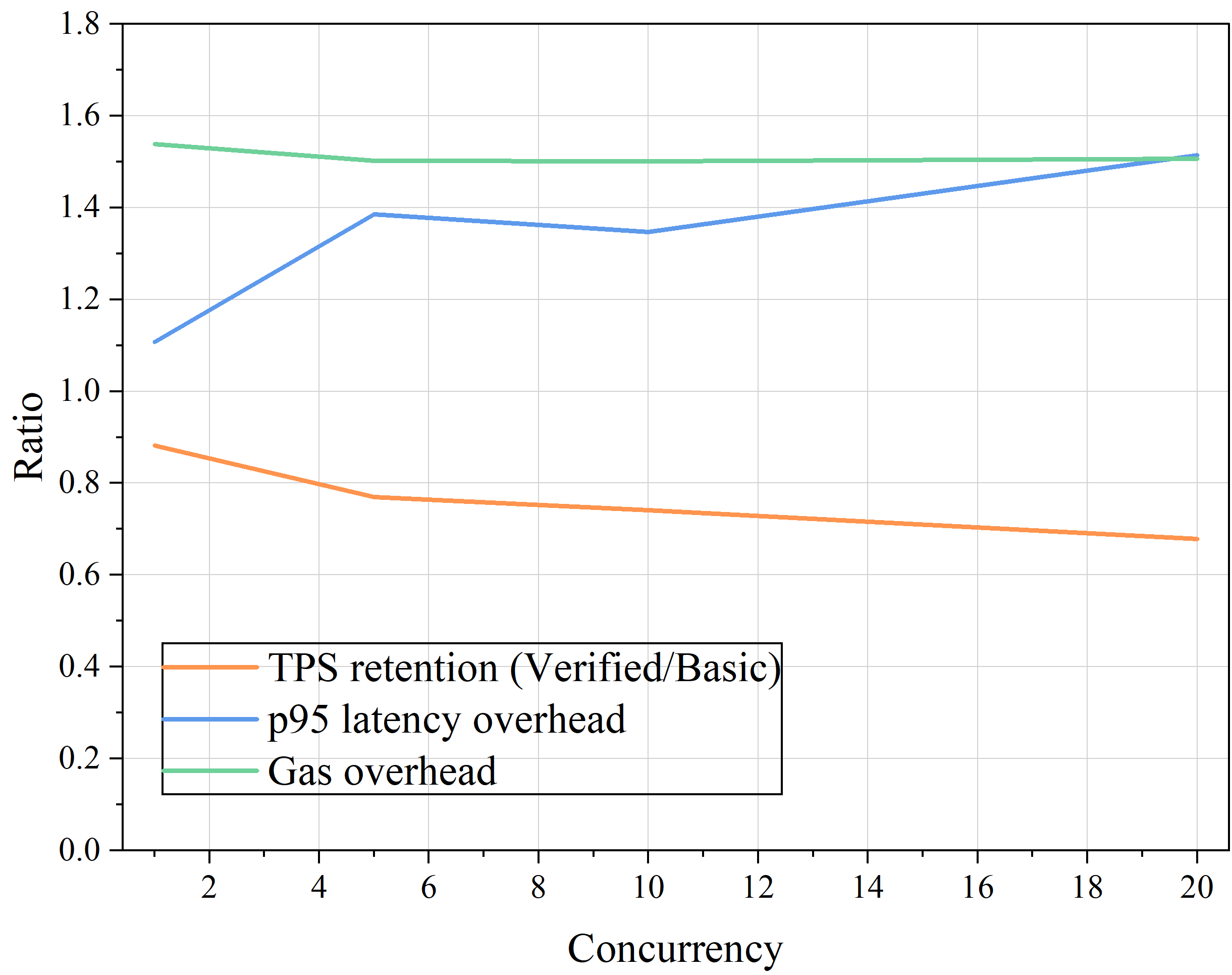}
\par\smallskip
{\scriptsize (d) TPS retention, p95 latency overhead,
and gas overhead ratios.\par}
\end{minipage}

\caption{Ganache smart-contract execution comparison after adding
verified-penalty logic.}
\label{fig:blockchain_performance}
\end{figure*}

\noindent
To verify that the contribution-credit logic can be executed on-chain, two complementary blockchain benchmarks are conducted. First, two Solidity implementations are deployed in a local Ganache EVM environment: a basic contribution-credit contract and a revised verified-penalty contract. The revised contract additionally records delayed verification, soft-indicator misreporting deviation, escrowed rewards, and penalty values. Each transaction corresponds to one incentive-relevant event. Ganache provides a reproducible contract-level benchmark for correctness, transaction latency, throughput, success rate, and gas consumption.

Second, the same contribution-credit workflow is implemented as Hyperledger Fabric chaincode and deployed on a Docker-based Fabric 2.5.15 test network with two peer organizations and a Raft orderer. The Fabric network is benchmarked through Hyperledger Caliper and a Fabric SDK timing script, which records per-transaction commit latency. This setting is closer to the proposed permissioned electricity-data consortium because endorsement, ordering, validation, and commit are all executed by a multi-organization network rather than a single local EVM node.

Blockchain execution efficiency is evaluated as a cost--capacity trade-off. Throughput measures the number of incentive-relevant events processed per second, commit latency measures execution timeliness, success rate verifies correct completion, and gas consumption represents the incremental EVM execution cost. Because delayed verification and report-linked penalties require additional state writes and arithmetic checks, the relevant implementation criterion is whether these functions remain executable with bounded overhead. The benchmark therefore quantifies the capacity and cost boundary of the proposed logic across Ganache and Hyperledger Fabric.

The Ganache benchmark results are summarized in Table~\ref{tab:ganache_performance}. Both contracts successfully executed all tested transactions. The verified-penalty contract introduces additional storage writes and arithmetic checks, so its throughput is lower and its gas consumption is higher than the basic contribution-credit contract. Under the largest tested workload, the revised contract reaches 17.36 TPS with p95 latency of 1286.41 ms and average gas consumption of 305799 gas, compared with 25.60 TPS, 849.52 ms, and 202973 gas for the basic contract. The results indicate that the added incentive-compatibility logic has a measurable but bounded execution cost in the local EVM benchmark.

\begin{table}[!htbp]
\centering
\caption{Ganache-based smart-contract execution benchmark.}
\label{tab:ganache_performance}
\scriptsize
\setlength{\tabcolsep}{3pt}
\renewcommand{\arraystretch}{1.08}
\begin{adjustbox}{max width=\columnwidth}
\begin{tabular}{lrrrrrrr}
\toprule
Contract & Tx & Conc. & Success & TPS & Mean & p95 & Gas \\
\midrule
Basic & 50 & 1 & 100\% & 7.99 & 124.7 & 141.7 & 217921 \\
Basic & 100 & 5 & 100\% & 19.63 & 253.9 & 279.5 & 200971 \\
Basic & 200 & 10 & 100\% & 23.54 & 423.5 & 480.4 & 201170 \\
Basic & 400 & 20 & 100\% & 25.60 & 779.1 & 849.5 & 202973 \\
Verified & 50 & 1 & 100\% & 7.05 & 141.6 & 156.8 & 335111 \\
Verified & 100 & 5 & 100\% & 15.12 & 329.8 & 387.3 & 301744 \\
Verified & 200 & 10 & 100\% & 17.45 & 571.8 & 647.0 & 301943 \\
Verified & 400 & 20 & 100\% & 17.36 & 1149.4 & 1286.4 & 305799 \\
\bottomrule
\end{tabular}
\end{adjustbox}
\end{table}

The Fabric benchmark is reported in Table~\ref{tab:fabric_performance}. Under target rates of 10, 25, and 50 TPS, the measured throughput reaches 10.05, 24.96, and 49.73 TPS, respectively, with a 100\% success rate in all three settings. Mean commit latency is 489.10 ms, 255.38 ms, and 166.12 ms, while p95 latency is 937 ms, 433 ms, and 270 ms. The lower latency under higher tested arrival rates is consistent with Fabric's block batching behavior: at low arrival rates, transactions wait longer for block formation, while higher rates fill blocks more quickly within this experimental range. A separate hot-key stress run, in which repeated transactions updated a small set of actor-credit keys, produced MVCC read conflicts; therefore, the reported benchmark uses unique actor-event keys to measure network execution capacity rather than intentional state-contention failure. Fig.~\ref{fig:blockchain_performance} focuses on the Ganache contract-level overhead introduced by the verified-penalty design, while Table~\ref{tab:fabric_performance} is retained as a permissioned-network capacity reference for the broader consortium deployment.

\begin{table}[!htbp]
\centering
\caption{Hyperledger Fabric permissioned-network benchmark.}
\label{tab:fabric_performance}
\scriptsize
\setlength{\tabcolsep}{2.4pt}
\renewcommand{\arraystretch}{1.08}
\begin{adjustbox}{max width=\columnwidth}
\begin{tabular}{lrrrrrrr}
\toprule
Load & Tx & Target & Success & TPS & Mean & p95 & p99 \\
\midrule
low & 100 & 10 & 100\% & 10.05 & 489.1 & 937 & 945 \\
medium & 300 & 25 & 100\% & 24.96 & 255.4 & 433 & 467 \\
high & 600 & 50 & 100\% & 49.73 & 166.1 & 270 & 294 \\
\bottomrule
\end{tabular}
\end{adjustbox}
\end{table}

\section{Conclusion}
\noindent
This paper develops a China-centered blockchain-enabled contribution-credit framework for electricity data assets. The main idea is to convert incentive-relevant electricity records into finalized contribution indicators, update dynamic credit through authenticated hard evidence and delayed soft-report settlement, and map the resulting signals into governance-oriented priority adjustment or mutual-aid settlement. The Chinese market is used as the main analytical setting because it combines hierarchical governance, interprovincial coordination, renewable accommodation, and cross-organizational support needs. The UK and US markets are retained as extensions because their institutional logic differs, but their incentive problems can still be represented through the same lower-layer contribution-credit structure.

The evaluation distinguishes three efficiency dimensions. In the behavioural simulation, the Proposed mechanism sharply reduces report inflation and achieves the highest transfer-neutral system surplus under the specified cost structure, while the Deviation penalty only regime yields slightly higher effort and participant utility. The rule-level workflow analysis shows how report registration, evidence verification, credit updating, and incentive execution are connected in a testable sequence without assigning a market-wide time or cost effect. The Ganache and Hyperledger Fabric experiments quantify implementation efficiency through throughput, latency, success rate, and gas consumption. The additional verified-penalty logic increases gas consumption and latency but remains executable with bounded overhead in the tested environments. These findings show that verification-aware incentive alignment can improve reporting compliance and distortion-adjusted resource outcomes while incurring explicit on-chain execution costs.

Several limitations remain. The China calibration in Appendix~C is an illustrative worked example rather than an empirical estimate based on actor-level dispatch and settlement records. The UK and US appendices establish institutional transferability but do not estimate cross-market treatment effects. Future work should incorporate more granular operational data, test privacy-preserving evidence submission, and evaluate larger permissioned deployments with richer adversarial data-quality scenarios. Even with these limitations, the proposed framework provides a clearer mechanism for turning electricity data from passive records into trusted incentive evidence.

\section*{Acknowledgment}
\noindent
This research was undertaken during Jiazheng Zhang's 12-month PhD research visit to the Financial Engineering Lab at the University of Bristol, under the supervision of Dr Wei Wei and Professor John Cartlidge, with support from the Jiangsu Province Higher Education Philosophy and Social Sciences Laboratory Project, the China Scholarship Council (grant no. 202506090237), and the UK Research and Innovation (UKRI) Engineering and Physical Sciences Research Council (EPSRC), grant no. EP/Y028392/1: AI for Collective Intelligence (AI4CI).

\appendix

\section{UK Institutional Extension: Capacity Commitment, Delivery, and Settlement}
\label{app:uk-extension}

\begin{figure*}[!t]
    \centering
    \includegraphics[width=0.98\textwidth]{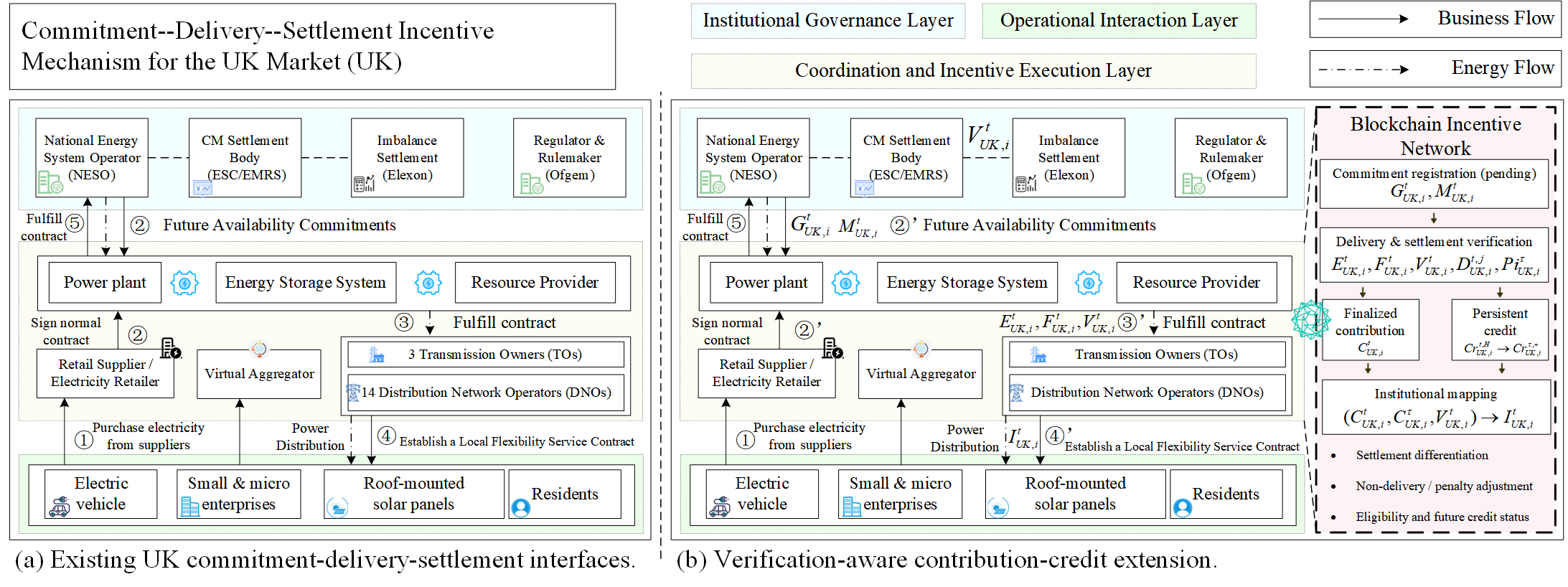}
    \caption{UK extension of the contribution-credit workflow. Subfigure (a) summarizes the existing commitment, delivery, and settlement interfaces; subfigure (b) adds authenticated record registration, delayed verification, credit updating, and rule-based incentive feedback. The diagram is an institutional mapping of the China-centered lower-layer logic, not a claim that existing UK settlement lacks performance verification.}
    \label{fig:uk}
\end{figure*}

\noindent
The UK extension addresses the separation between forward capacity commitments and evidence of delivery during stress or balancing events. Capacity providers register availability and commitment information before delivery, while NESO, the Capacity Market settlement arrangements, Elexon, network operators, and metering systems produce different parts of the subsequent evidence chain. Existing UK arrangements already include prequalification, testing, satisfactory-performance requirements, non-delivery penalties, and settlement procedures. The proposed extension does not replace them. It adds an auditable link between the original commitment, the later delivery evidence, the resulting credit state, and the applicable settlement consequence.

Table~\ref{tab:extension-indicator-mapping} records how the indicator roles used in the China-centered mechanism are reinterpreted in the two institutional extensions. The mapping transfers the functional sequence, not the China-specific coefficients or governance outputs.

\begin{table*}[!t]
\centering
\caption{Indicator-role Mapping for the UK and US Extensions}
\label{tab:extension-indicator-mapping}
\scriptsize
\setlength{\tabcolsep}{4pt}
\renewcommand{\arraystretch}{1.08}
\begin{tabularx}{\textwidth}{p{0.11\textwidth}XX}
\toprule
Role & UK extension & US extension \\
\midrule
$G$ & Registered availability or capacity commitment & Registered large-load, flexibility, or electricity-attribute commitment \\
$E$ & Metered delivery of committed capacity or flexibility & Metered load response or executed flexibility \\
$M$ & Improvement in delivery, imbalance, or settlement performance & Improvement in planning support, flexibility accountability, or attribute matching \\
$F$ & Delivery or availability-fulfilment quality & Response or obligation-fulfilment quality \\
$V$ & Non-delivery, imbalance, or settlement inconsistency & Disclosure, response, interconnection, or attribute-claim inconsistency \\
$I$ & Settlement, penalty, eligibility, or credit differentiation & Flexibility reward, accountability adjustment, or future participation condition \\
\bottomrule
\end{tabularx}
\end{table*}

The UK extension reuses the common equations with UK-indexed inputs. The indicator roles reported in Table~10 are normalized according to Eq.~(1), using bounds fixed before the corresponding commitment is submitted. Equation~(2) calculates finalized contribution, while Eqs.~(3a) and~(3b) update credit from authenticated hard evidence and delayed soft-report settlement. Availability declarations and improvement claims remain in the soft-report set $\mathcal{S}_{\mathrm{UK}}$ until the required delivery or settlement reference becomes available. Metered delivery, authenticated non-delivery and imbalance records, and finalized settlement data form the hard-evidence set $\mathcal{H}_{\mathrm{UK}}$. Report deviation and the realized report-linked penalty follow Eqs.~(11) and~(12), with the tolerance, audit probability, penalty intensity, and verification delay fixed by the competent UK institutions.

The UK-specific difference lies in the upper-layer institutional mapping. Relative to the governance-oriented mapping in Eq.~(4), the UK output includes the inconsistency deduction $-\lambda_{\mathrm{UK},V}\widetilde{V}_{\mathrm{UK},i}^{t}$ and maps the resulting score to settlement, eligibility, penalty, or reward differentiation instead of Chinese governance priority. Higher finalized contribution and verified credit increase the applicable score, whereas non-delivery, imbalance, or settlement inconsistency reduces it. Under the same assumptions, Proposition 1 remains applicable after substituting the UK evidence law and enforcement parameters. No China-specific numerical weight is transferred to the UK extension.


\section{US Institutional Extension: Large-load Disclosure and Flexibility Accountability}
\label{app:us-extension}

\begin{figure*}[!t]
    \centering
    \includegraphics[width=0.98\textwidth]{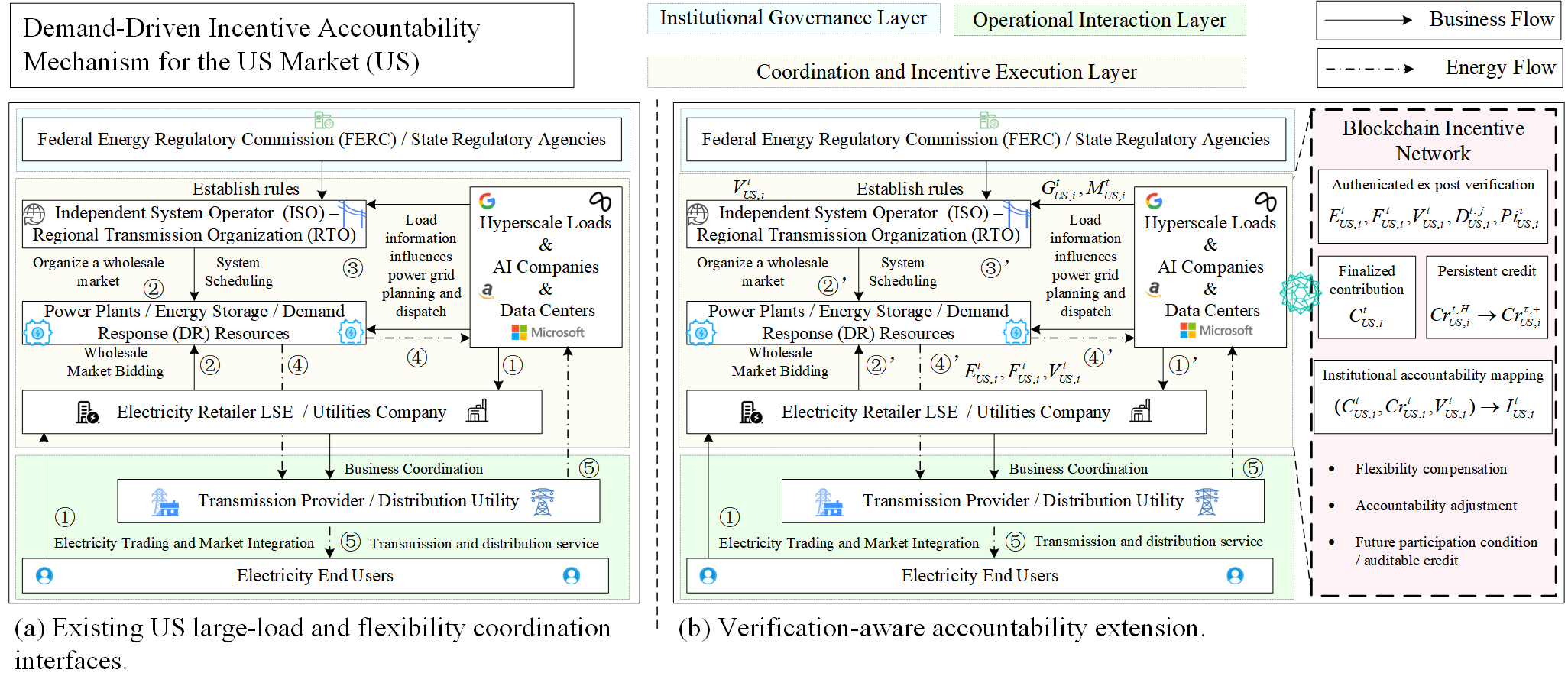}
    \caption{US extension of the contribution-credit workflow. Subfigure (a) summarizes the existing market and network interfaces; subfigure (b) links registered large-load or flexibility commitments to authenticated execution evidence, delayed verification, credit updating, and accountability consequences.}
    \label{fig:us}
\end{figure*}

\noindent
The US extension addresses information and accountability problems created by rapid large-load growth, heterogeneous ISO/RTO arrangements, and increasingly granular flexibility and electricity-attribute claims. Large-load participants may possess private information about the scale, timing, and location of future demand, while metered demand, interconnection milestones, executed response, and certificate settlement become observable through different institutional systems. The proposed layer records the commitment and its evidence pointer when submitted, but leaves the claim provisional until an authorized reference is available.

The US extension applies the same normalization, contribution, credit-updating, and delayed-verification rules with US-indexed inputs. Future large-load disclosures, flexibility commitments, and favourable electricity-attribute claims remain in $\mathcal{S}_{\mathrm{US}}$ until an authorised reference becomes available. Metered demand, executed response, authenticated interconnection milestones, and certificate or settlement records form $\mathcal{H}_{\mathrm{US}}$. Their normalized values enter Eq.~(2), hard evidence updates credit through Eq.~(3a), and due soft reports are settled through Eq.~(3b). Report deviation and the realised penalty follow Eqs.~(11) and~(12) using US-specific verification and enforcement parameters.

The market-specific difference again lies in the upper-layer output. Relative to Eq.~(4), the US mapping includes the inconsistency deduction $-\lambda_{\mathrm{US},V}\widetilde{V}_{\mathrm{US},i}^{t}$, but the resulting output is interpreted as flexibility compensation, an accountability adjustment, a future participation condition, or an auditable credit record. These consequences differ from both Chinese governance priority and UK capacity-settlement differentiation. Because institutional authority is distributed across ISO/RTO, utility, interconnection, and electricity-attribute arrangements, the coefficients must be authorised for the relevant jurisdiction; neither Chinese nor UK numerical weights are transferred.


\section{Illustrative Calibration and Worked Examples for the China Mechanism}
\label{app:china-calibration}

\begin{table*}[!t]
\centering
\caption{Illustrative AHP--Entropy Weights for the China Worked Examples}
\label{tab:china-hybrid-weights}

\scriptsize
\setlength{\tabcolsep}{8pt}
\renewcommand{\arraystretch}{1.08}

\begin{tabular}{@{} l c l l @{}}
\toprule
Indicator group & Order & Weight source & Weight vector \\
\midrule

Scene~1 contribution
& $(G,E,M,F)$
& AHP
& $(0.2437,0.4379,0.0994,0.2190)$ \\

& & Entropy
& $(0.1963,0.2274,0.2118,0.3645)$ \\

& & Hybrid ($\theta=0.7$)
& $(0.2295,0.3748,0.1331,0.2626)$ \\

\addlinespace

Scene~1 credit
& $(E,M,F,V)$
& AHP
& $(0.4673,0.0954,0.2772,0.1601)$ \\

& & Entropy
& $(0.2660,0.2477,0.4264,0.0599)$ \\

& & Hybrid ($\theta=0.7$)
& $(0.4069,0.1411,0.3219,0.1300)$ \\

\addlinespace

Scene~2 contribution
& $(H,Q,U,B_G)$
& AHP
& $(0.2772,0.4673,0.1601,0.0954)$ \\

& & Entropy
& $(0.2549,0.5345,0.0234,0.1872)$ \\

& & Hybrid ($\theta=0.7$)
& $(0.2705,0.4874,0.1191,0.1230)$ \\

\addlinespace

Scene~2 bilateral credit
& $(Q,U,B)$
& AHP
& $(0.5396,0.2970,0.1634)$ \\

& & Entropy
& $(0.7173,0.0314,0.2513)$ \\

& & Hybrid ($\theta=0.7$)
& $(0.5929,0.2173,0.1898)$ \\

\bottomrule
\end{tabular}

\end{table*}

\subsection{Scope of the Calibration}
\noindent
This appendix illustrates how weights can be selected and how the China equations operate once normalized evidence has been supplied. It is not an empirical estimate of provincial behaviour and does not infer mechanism effectiveness from aggregate national statistics. The official quantities in Table~\ref{tab:china-public-anchors} establish only the scale and relevance of renewable accommodation and interregional coordination. Actor-level commitments, delayed verification records, bilateral deviations, and mutual-aid settlements are not publicly observed in a common dataset.

For an indicator group with AHP weight $w_j^{\mathrm{AHP}}$ and entropy weight $w_j^{\mathrm{ENT}}$, the illustrative hybrid rule is
\begin{equation}
w_j=\theta w_j^{\mathrm{AHP}}+(1-\theta)w_j^{\mathrm{ENT}},
\qquad 0\leq\theta\leq1.
\label{eq:hybrid-weight}
\end{equation}
AHP represents an ex ante policy scorecard, while the entropy term measures dispersion in an illustrative calibration matrix. Neither component is estimated from the national totals. The worked example uses $\theta=0.7$; alternative values should be reported whenever the weights are used in a substantive application.

The four AHP pairwise-comparison matrices used for the contribution, Scene~1 credit, event contribution, and bilateral-credit indicator groups are
\begin{equation}
\resizebox{\columnwidth}{!}{$\displaystyle
A_C=\left[\begin{smallmatrix}
1&1/2&3&1\\
2&1&4&2\\
1/3&1/4&1&1/2\\
1&1/2&2&1
\end{smallmatrix}\right],
\quad
A_{Cr}=\left[\begin{smallmatrix}
1&4&2&3\\
1/4&1&1/3&1/2\\
1/2&3&1&2\\
1/3&2&1/2&1
\end{smallmatrix}\right]
$}
\label{eq:china-ahp-scene1}
\end{equation}

\begin{equation}
\resizebox{\columnwidth}{!}{$\displaystyle
A_k=\left[\begin{smallmatrix}
1&1/2&2&3\\
2&1&3&4\\
1/2&1/3&1&2\\
1/3&1/4&1/2&1
\end{smallmatrix}\right],
\quad
A_{Bi}=\left[\begin{smallmatrix}
1&2&3\\
1/2&1&2\\
1/3&1/2&1
\end{smallmatrix}\right]
$}
\label{eq:china-ahp-scene2}
\end{equation}
Their consistency ratios are 0.003, 0.005, 0.003, and 0.000, respectively. Table~\ref{tab:china-hybrid-weights} reports the resulting illustrative wei\-ghts. Before the entropy calculation, adverse deviation dimensions are benefit-oriented as $1-V$ or $1-B$; the analytical equations retain $V$ and $B$ as penalty terms. The entropy vectors are dispersion weights for the worked calibration profiles; they are not market-estimated coefficients.

For Eq.~\eqref{eq:event_contribution}, the first three benefit weights in the Scene~2 contribution row are renormalized to satisfy $\gamma_H+\gamma_Q+\gamma_U=1$, giving $(\gamma_H,\gamma_Q,\gamma_U)=(0.3084, 0.5558,\allowbreak 0.1358)$; the deviation coefficient remains $\gamma_B=0.1230$. For the Scene~1 example, the contribution weights are used directly. Credit uses $\rho=0.60$, $(\beta_E,\beta_M,\beta_F,\beta_V)=(0.4069,\allowbreak 0.1411,\allowbreak 0.3219,\allowbreak 0.1300)$, and $\beta_G=0$ in the persistent-credit update. Thus, $G$ affects finalized contribution but not persistent credit in this particular calibration. This choice is illustrative and must be fixed before reports are submitted.

\subsection{Scene 1 Worked Calculation}
\noindent
Table~\ref{tab:china-scene1-inputs} gives one normalized period for a sending and a receiving entity. These are calculation inputs, not observed provincial records and not ``before'' and ``after'' mechanism outcomes. Beginning credit is $Cr_i^t=0.50$, and the incentive mapping uses $(\lambda_C,\lambda_R)=(0.75,0.25)$.

\begin{table}[!t]
\centering
\caption{Illustrative Normalized Inputs for the Scene~1 Calculation}
\label{tab:china-scene1-inputs}
\scriptsize
\setlength{\tabcolsep}{3pt}
\renewcommand{\arraystretch}{1.08}
\begin{tabular}{lccccc}
\toprule
Entity & $\widetilde G$ & $\widetilde E$ & $\widetilde M$ & $\widetilde F$ & $\widetilde V$ \\
\midrule
Sending & 0.82 & 0.79 & 0.74 & 0.83 & 0.05 \\
Receiving & 0.76 & 0.81 & 0.75 & 0.84 & 0.04 \\
\bottomrule
\end{tabular}
\end{table}

Assume first that both soft $M$ reports are audited and accepted, so $A_i^{t,M}=1$ and $\Pi_i^{t+\ell_M}=0$. Applying Eqs.~\eqref{eq:contribution_score}, \eqref{eq:hard_credit_update}, \eqref{eq:soft_credit_update}, and \eqref{eq:period_incentive} yields Table~\ref{tab:china-scene1-results}. The current incentive uses beginning credit, as required by Eq.~\eqref{eq:period_incentive}; the post-settlement credit affects future periods.

\begin{table}[!t]
\centering
\caption{Outputs of the Scene~1 Worked Calculation}
\label{tab:china-scene1-results}
\scriptsize
\setlength{\tabcolsep}{3pt}
\renewcommand{\arraystretch}{1.08}
\begin{tabular}{lcccc}
\toprule
Entity & $C_i^t$ & $Cr_i^{t,H}$ & $Cr_i^{t+\ell_M,+}$ & $I_i^t$ \\
\midrule
Sending & 0.8007 & 0.5832 & 0.5920 & 0.7256 \\
Receiving & 0.7984 & 0.5890 & 0.5981 & 0.7238 \\
\bottomrule
\end{tabular}
\end{table}

To illustrate the report-linked consequence, suppose the sending entity instead reports $\widehat M=0.84$, the authenticated reference is $z=0.74$, $\varepsilon_M=0.01$, $A=1$, and $\pi_M=0.30$. Then $D=0.09$ and $\Pi=0.027$. With $\beta_P=0.50$, post-settlement credit becomes 0.5785 rather than 0.5920, in addition to the applicable escrow forfeiture. This is a mechanical illustration of the rule, not an estimate of actual misreporting.

\subsection{Scene 2 Worked Calculation}

\noindent
Consider a registered mutual-aid event with $H_k=1500$ MWh, $P_k=500$ RMB/MWh, $(\widetilde H_k,\widetilde Q_k,\widetilde U_k)=(0.88,\allowbreak 0.91,\allowbreak 0.85)$, $\widetilde B_{G,k}=0.05$, $\widetilde B_{R,k}=0.04$, and $\widetilde{\mathrm{Pay}}_k=0.93$. Let pre-event credits be $Cr_{G,k}^{-}=0.55$ and $Cr_{R,k}^{-}=0.53$, and let no contractual or report-linked penalty be due. The bilateral-credit coefficients are $(\zeta_Q,\zeta_U,\zeta_B)=(0.5929,0.2173,0.1898)$.

The event equations then give
\begin{equation}
\begin{aligned}
C_k&=0.8761, Cr_{G,k}^{+}=0.6453, Cr_{R,k}^{+}=0.6824.
\end{aligned}
\end{equation}
With $\mu_C=100000$~RMB and $\mu_R=50000$~RMB, Eq.~\eqref{eq:event_settlement} gives
\begin{equation}
\begin{aligned}
S_k={}&500(1500)+100000(0.8761)+50000(0.55)\\
={}&865106\ \mathrm{RMB},
\end{aligned}
\end{equation}
up to rounding. The settlement uses pre-event supporting credit, so the current event is not rewarded twice through its own credit update. The example shows the calculation order only; it is not a recommended tariff or a measured settlement value.

\section{Cross-market Public-data Context and Limitations}
\label{app:cross-market-context}
\begin{figure*}[!t]
\centering
\begin{minipage}[t]{0.235\textwidth}
\centering
\includegraphics[width=\linewidth]{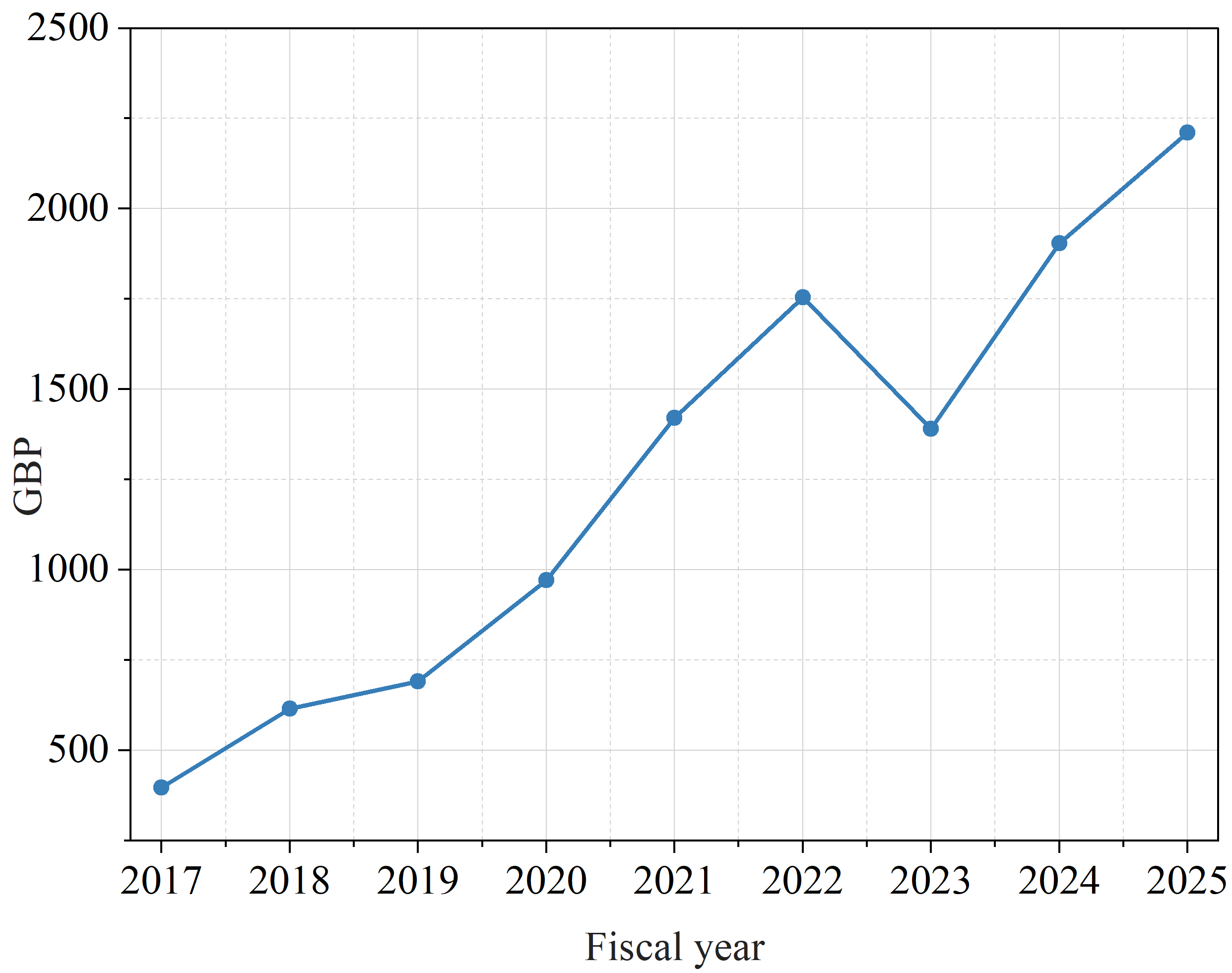}
\par{\scriptsize (a) UK annual constraint costs (NESO).}
\end{minipage}\hfill
\begin{minipage}[t]{0.235\textwidth}
\centering
\includegraphics[width=\linewidth]{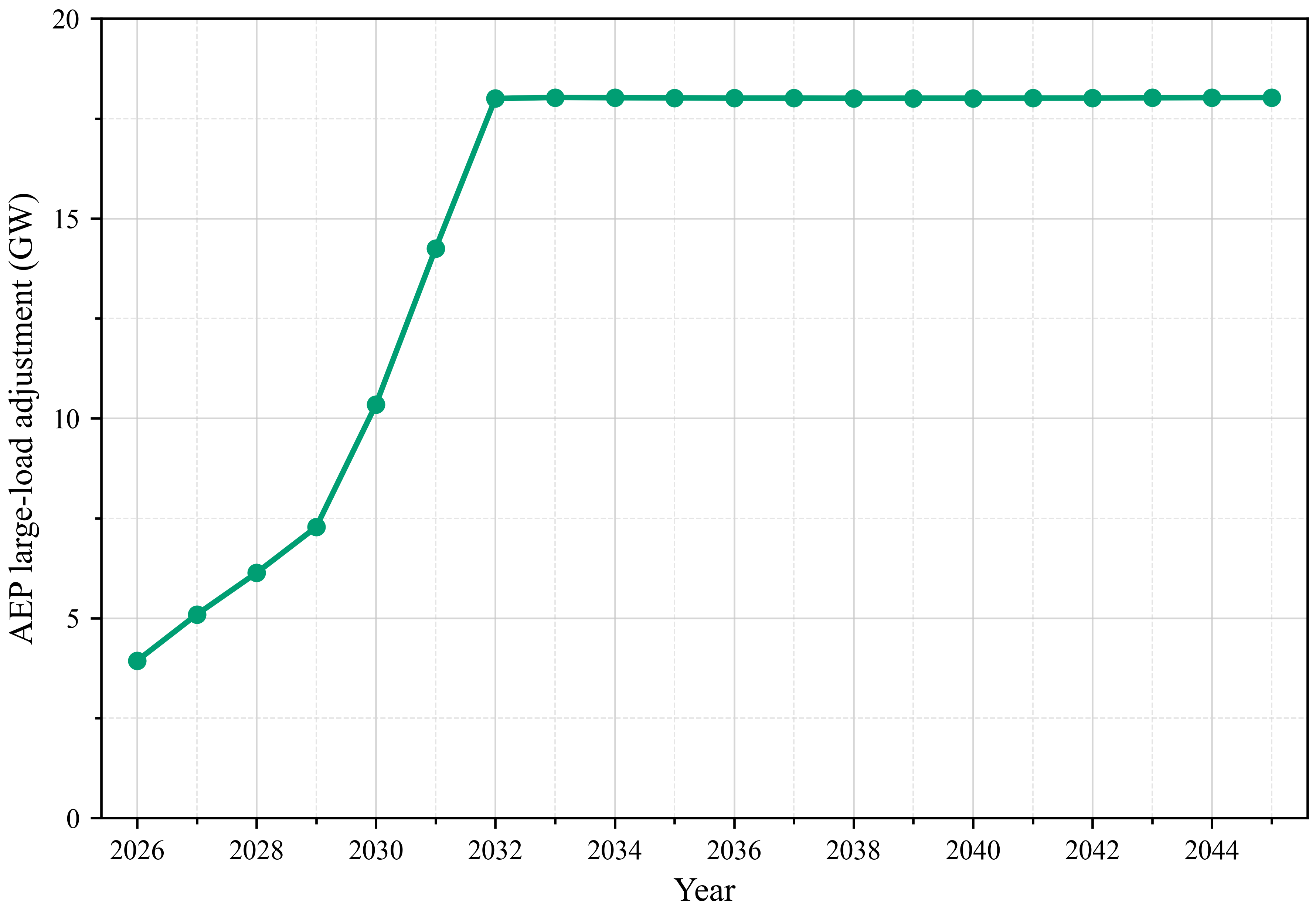}
\par{\scriptsize (b) AEP-zone large-load adjustment,
2026--2045 (PJM Table B-9b \cite{pjm2025loadforecast}).}
\end{minipage}\hfill
\begin{minipage}[t]{0.235\textwidth}
\centering
\includegraphics[width=\linewidth]{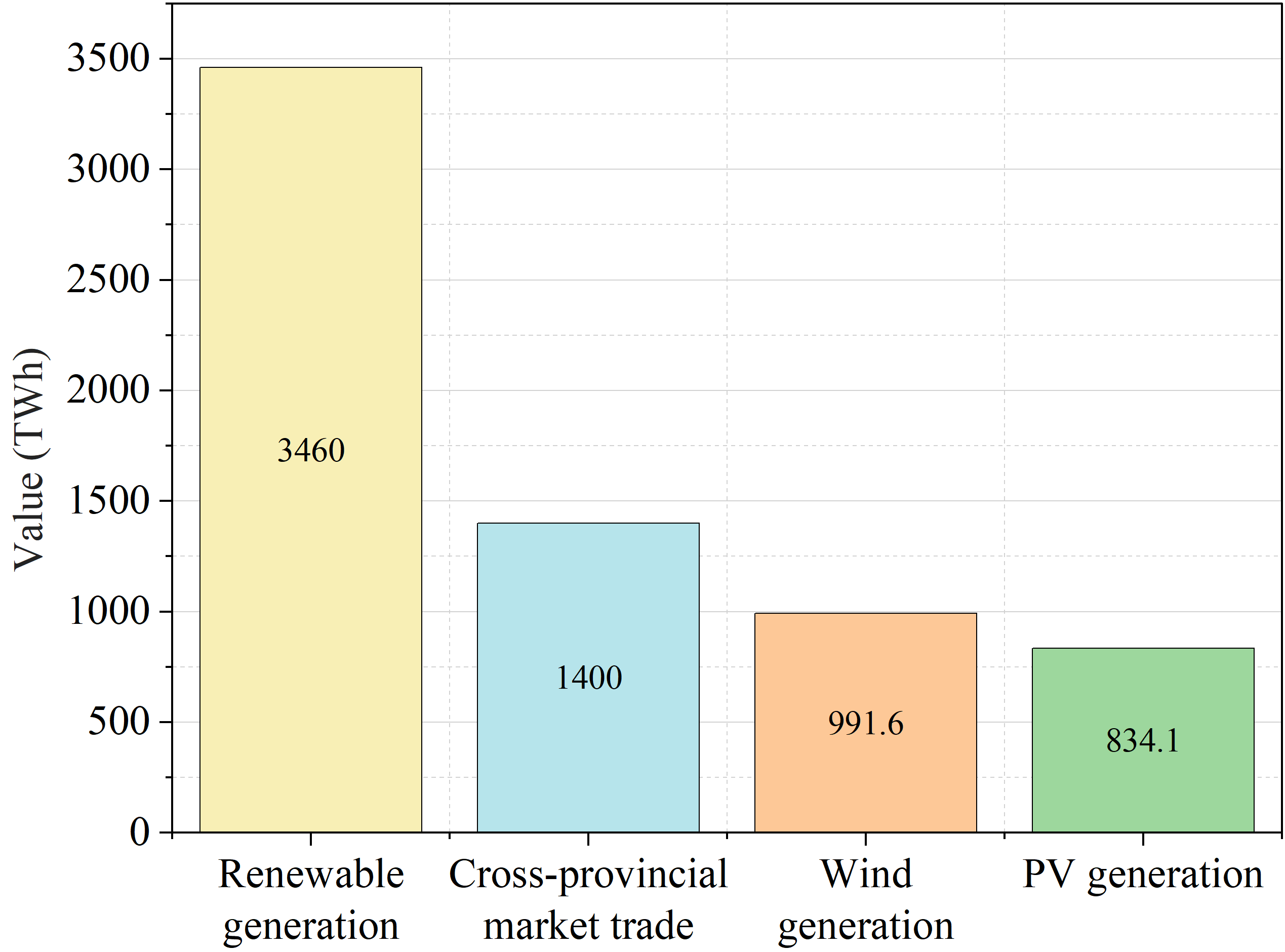}
\par{\scriptsize (c) China renewable and trading scale (NEA).}
\end{minipage}\hfill
\begin{minipage}[t]{0.235\textwidth}
\centering
\includegraphics[width=\linewidth]{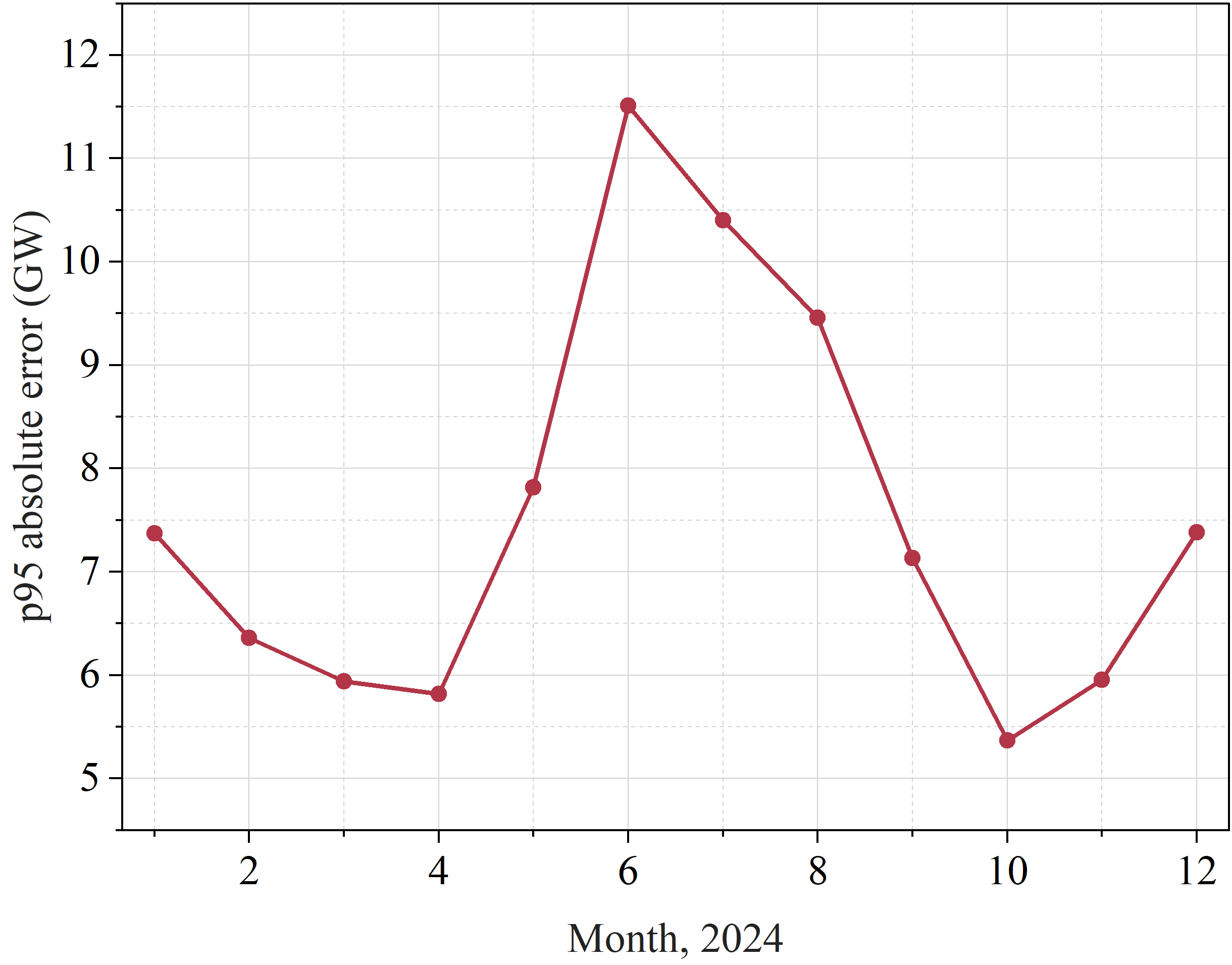}
\par{\scriptsize (d) Monthly p95 PJM day-ahead
forecast error (EIA-930).}
\end{minipage}
\caption{Cross-market public-data context. The panels describe
operating conditions, not effects of the proposed mechanism.}
\label{fig:cross-market-public-context}
\end{figure*}

\begin{table}[!tbp]
\centering
\caption{Public-data Context for the Cross-market Extensions}
\label{tab:cross-market-public-anchors}
\scriptsize
\setlength{\tabcolsep}{3pt}
\renewcommand{\arraystretch}{1.08}
\begin{tabularx}{\columnwidth}{
  >{\raggedright\arraybackslash}p{0.10\columnwidth}Y}
\toprule
Market & Public-data context and limitation \\
\midrule
China &
\textit{Source:} NEA renewable-operation and electricity-market statistics.
\textit{Context:} 3.46~PWh renewable generation; 95.9\% wind and
96.8\% PV utilization; 1.40~PWh interregional market-based trading
in 2024.
\textit{Use:} Establishes the scale of the main Chinese scenarios;
no actor-level contribution or settlement effect is inferred. \\
\addlinespace[2pt]
UK &
\textit{Source:} NESO operational datasets and EMRS Capacity Market
figures \cite{neso2026dataportal,emrs2026keyfigures}.
\textit{Context:} GBP~1.26~billion in annual Capacity Market payments
and 10.88~TWh of gross demand during high-demand periods for the
2024/25 delivery year.
\textit{Use:} Establishes commitment and settlement exposure;
aggregate figures do not identify a verification-rule effect. \\
\addlinespace[2pt]
US &
\textit{Source:} EIA-930 PJM demand data; the 2025 PJM Long-Term Load
Forecast Report and accompanying Table B-9b; and FERC Docket RM26-4
\cite{eia2026opendata,pjm2025loadforecast,ferc2026rm264}.
\textit{Context:} 8,736 matched hourly observations in 2024;
153.1~GW peak demand; 8.1~GW p95 absolute day-ahead forecast error;
and the AEP large-load adjustment trajectory for 2026--2045.
\textit{Use:} Establishes forecast and disclosure pressure; regional
observations do not reveal participant-level choices. \\
\bottomrule
\end{tabularx}
\end{table}

\noindent
The UK and US extensions are retained to demonstrate institutional transferability, but they are not co-equal empirical cases in the main text. Table~\ref{tab:cross-market-public-anchors} records the public quantities used to describe their operating context. The values establish orders of magnitude and data availability; they do not estimate the effect of the proposed rules.

The EIA-930 quantities are author calculations after aligning hourly actual-demand and day-ahead forecast series and removing unmatched observations; p95 denotes the 95th percentile of the absolute forecast error. Figure~\ref{fig:cross-market-public-context} preserves the original cross-market public-data panels as contextual material in the appendix. It is not evidence of behavioural improvement or of a common causal effect across the three markets.

Panel~(b) of Fig.~\ref{fig:cross-market-public-context} is an author reconstruction of the AEP large-load adjustment series reported in Table~B-9b accompanying the 2025 PJM Long-Term Load Forecast Report \cite{pjm2025loadforecast}. PJM attributes the AEP adjustment to growth in data-center load and a chip-processing facility. The series is used only to describe the scale and timing of forecast adjustments; it does not represent realized load, participant behaviour, or an effect of the proposed mechanism.

\clearpage
\balance
\bibliographystyle{elsarticle-num}
\bibliography{refs}

@article{gough2022privacy,
author  = {Gough, M. and Santos, S. and Alskaif, T. and Javadi, M. and Castro, R. and Catalao, J. P. S.},
title   = {Preserving Privacy of Smart Meter Data in a Smart Grid Environment},
journal = {IEEE Transactions on Industrial Informatics},
volume  = {18},
number  = {1},
pages   = {707--718},
year    = {2022}
}

@article{mitra2024smartmeteranalytics,
author  = {Mitra, S. and Chakraborty, B. and Mitra, P.},
title   = {Smart Meter Data Analytics Applications for Secure, Reliable and Robust Grid System: Survey and Future Directions},
journal = {Energy},
volume  = {289},
pages   = {129920},
year    = {2024}
}

@article{cabot2024demand,
author  = {Cabot, C. and Villavicencio, M.},
title   = {The Demand-side Flexibility in Liberalised Power Market: A Review of Current Market Design and Objectives},
journal = {Renewable and Sustainable Energy Reviews},
volume  = {201},
pages   = {114643},
year    = {2024}
}

@article{morales2022demand,
author  = {Morales-Espana, G. and Martinez-Gordon, R. and Sijm, J.},
title   = {Classifying and Modelling Demand Response in Power Systems},
journal = {Energy},
volume  = {242},
pages   = {122544},
year    = {2022}
}

@article{zhu2025marketdesign,
author  = {Zhu, Z. and Bu, S. and Chan, K. W. and Li, F. and Ye, Y. and Chung, C. Y. and Strbac, G.},
title   = {Designing the Future Electricity Spot Market with High Renewables via Reliable Simulations},
journal = {Nature Reviews Electrical Engineering},
volume  = {2},
pages   = {320--337},
year    = {2025}
}

@article{mazrae2025transactive,
author  = {Mazrae, A. K. and Naderian, H. and Baghaee, H. R. and Sheikh-El-Eslami, M. K. and Karimi, M.},
title   = {Transactive Energy and Peer-to-peer Energy Trading Based on Blockchain: A Comprehensive Review and a Generalized Cyber-physical Framework},
journal = {Energy Strategy Reviews},
volume  = {62},
pages   = {101949},
year    = {2025}
}

@article{veldkamp2023data,
author  = {Veldkamp, L.},
title   = {Valuing Data as an Asset},
journal = {Review of Finance},
volume  = {27},
number  = {5},
pages   = {1545--1562},
year    = {2023}
}

@article{jones2020data,
author  = {Jones, C. I. and Tonetti, C.},
title   = {Nonrivalry and the Economics of Data},
journal = {American Economic Review},
volume  = {110},
number  = {9},
pages   = {2819--2858},
year    = {2020}
}

@book{iea2023china,
author    = {{International Energy Agency}},
title     = {Building a Unified National Power Market System in China},
publisher = {OECD Publishing},
address   = {Paris, France},
year      = {2023},
doi       = {10.1787/8d4cc527-en}
}

@article{xiang2023china,
author  = {Xiang, C. and Zheng, X. and Song, F. and Lin, J. and Jiang, Z.},
title   = {Assessing the Roles of Efficient Market versus Regulatory Capture in China's Power Market Reform},
journal = {Nature Energy},
volume  = {8},
pages   = {747--757},
year    = {2023}
}

@article{yang2024greenpower,
author  = {Yang, Y.-S. and Xie, B.-C. and Tan, X.},
title   = {Impact of Green Power Trading Mechanism on Power Generation and Interregional Transmission in China},
journal = {Energy Policy},
volume  = {189},
pages   = {114088},
year    = {2024}
}

@article{wang2023uncertainty,
author  = {Wang, J. and Chen, L. and Tan, Z. and Du, E. and Liu, N. and Ma, J. and Sun, M. and Li, C. and Song, J. and Lu, X. and Tan, C.-W. and He, G.},
title   = {Inherent Spatiotemporal Uncertainty of Renewable Power in China},
journal = {Nature Communications},
volume  = {14},
pages   = {5379},
year    = {2023}
}

@article{yan2024crossprovincial,
author  = {Yan, S. and Wang, W. and Li, X. and He, H. and Zhao, X.},
title   = {Research on Cross-provincial Power Trading Strategy Considering the Medium and Long-term Trading Plan},
journal = {Scientific Reports},
volume  = {14},
pages   = {30137},
year    = {2024}
}

@article{tao2024capacity,
author  = {Tao, J. and Li, R. and Chen, S. and Qu, H. and Xiang, Y.},
title   = {Reverse Impact of Capacity Markets for a Renewable-dominated Power System},
journal = {Applied Energy},
volume  = {375},
pages   = {124154},
year    = {2024}
}

@article{zhao2024weather,
author  = {Zhao, J. and Li, F. and Zhang, Q.},
title   = {Impacts of Renewable Energy Resources on the Weather Vulnerability of Power Systems},
journal = {Nature Energy},
volume  = {9},
number  = {11},
pages   = {1407--1414},
year    = {2024}
}

@misc{desnz2024capacity,
author       = {{Department for Energy Security and Net Zero}},
title        = {Capacity Market: 10-year Review (2019 to 2024)},
year         = {2024},
address      = {London, U.K.},
howpublished = {\url{https://www.gov.uk/government/publications/capacity-market-10-year-review-2019-to-2024}},
note         = {Online; accessed 9 June 2026}
}

@article{masanet2020datacenter,
author  = {Masanet, E. and Shehabi, A. and Lei, N. and Smith, S. and Koomey, J.},
title   = {Recalibrating Global Data Center Energy-use Estimates},
journal = {Science},
volume  = {367},
number  = {6481},
pages   = {984--986},
year    = {2020}
}

@misc{iea2025energyai,
author       = {{International Energy Agency}},
title        = {Energy and AI},
year         = {2025},
address      = {Paris, France},
howpublished = {\url{https://www.iea.org/reports/energy-and-ai}},
note         = {Online; accessed 9 June 2026}
}

@article{colangelo2026ai,
author  = {Colangelo, P. and Coskun, A. K. and Megrue, J. and Roberts, C. and Sengupta, S. and Sivaram, V. and Tiao, E. and Vijaykar, A. and Williams, C. and Wilson, D. C. and others},
title   = {AI Data Centres as Grid-interactive Assets},
journal = {Nature Energy},
volume  = {11},
pages   = {254--261},
year    = {2026}
}

@article{bjorn2022certificates,
author  = {Bj{\o}rn, A. and Lloyd, S. M. and Brander, M. and Matthews, H. D.},
title   = {Renewable Energy Certificates Allow Companies to Overstate Their Emission Reductions},
journal = {Nature Climate Change},
volume  = {12},
number  = {6},
pages   = {508--509},
year    = {2022}
}

@article{xu2024system,
author  = {Xu, Q. and Ricks, W. and Manocha, A. and Patankar, N. and Jenkins, J. D.},
title   = {System-level Impacts of Voluntary Carbon-free Electricity Procurement Strategies},
journal = {Joule},
volume  = {8},
number  = {2},
pages   = {374--400},
year    = {2024}
}

@article{riepin2024means,
author  = {Riepin, I. and Brown, T.},
title   = {On the Means, Costs, and System-level Impacts of 24/7 Carbon-free Energy Procurement},
journal = {Energy Strategy Reviews},
volume  = {54},
pages   = {101488},
year    = {2024}
}

@article{scholta2025temporal,
author  = {Scholta, H. F. and Blaschke, M. J.},
title   = {Temporal Matching as an Accounting Principle for Green Electricity Claims},
journal = {Nature Communications},
volume  = {16},
pages   = {9280},
year    = {2025}
}

@article{egunjobi2024blockchain,
author  = {Egunjobi, O. O. and Gomes, A. and Egwim, C. N. and Morais, H.},
title   = {A Systematic Review of Blockchain for Energy Applications},
journal = {e-Prime - Advances in Electrical Engineering, Electronics and Energy},
volume  = {9},
pages   = {100751},
year    = {2024}
}

@article{he2024review,
author  = {He, Y. and Zhou, Z. and Pan, Y. and Chong, F. and Wu, B. and Xiao, K. and Li, H.},
title   = {Review of Data Security within Energy Blockchain: A Comprehensive Analysis of Storage, Management, and Utilization},
journal = {High-Confidence Computing},
volume  = {4},
number  = {3},
pages   = {100233},
year    = {2024}
}

@article{kirli2022smartcontracts,
author  = {Kirli, D. and Couraud, B. and Robu, V. and Salgado-Bravo, M. and Norbu, S. and Andoni, M. and Antonopoulos, I. and Negrete-Pincetic, M. and Flynn, D. and Kiprakis, A.},
title   = {Smart Contracts in Energy Systems: A Systematic Review of Fundamental Approaches and Implementations},
journal = {Renewable and Sustainable Energy Reviews},
volume  = {158},
pages   = {112013},
year    = {2022}
}

@article{junaidi2023blockchaindr,
author  = {Junaidi, N. and Abdullah, M. P. and Alharbi, B. and Shaaban, M.},
title   = {Blockchain-based Management of Demand Response in Electric Energy Grids: A Systematic Review},
journal = {Energy Reports},
volume  = {9},
pages   = {5075--5100},
year    = {2023}
}

@article{hassan2023oracle,
author  = {Hassan, A. and Makhdoom, I. and Iqbal, W. and Ahmad, A. and Raza, A.},
title   = {From Trust to Truth: Advancements in Mitigating the Blockchain Oracle Problem},
journal = {Journal of Network and Computer Applications},
volume  = {217},
pages   = {103672},
year    = {2023}
}

@misc{neso2026dataportal,
author       = {{National Energy System Operator}},
title        = {NESO Data Portal},
year         = {2026},
howpublished = {\url{https://www.neso.energy/data-portal}},
note         = {Online; accessed 9 June 2026}
}

@misc{eia2026opendata,
author       = {{U.S. Energy Information Administration}},
title        = {Open Data: Electricity},
year         = {2026},
howpublished = {\url{https://www.eia.gov/opendata/index.php/browser/browser/electricity/browser/}},
note         = {Online; accessed 9 June 2026}
}

@misc{emrs2026keyfigures,
author       = {{Electricity Market Reform Settlement Ltd.}},
title        = {Key Figures for Payments 2024--25},
year         = {2026},
address      = {London, U.K.},
howpublished = {\url{https://www.emrsettlement.co.uk/document/settlement-data/key-figures-for-payments-2024-25/}},
note         = {Online; accessed 9 June 2026}
}

@misc{ferc2026rm264,
author       = {{Federal Energy Regulatory Commission}},
title        = {Interconnection of Large Loads to the Interstate Transmission System, Docket No. RM26-4-000},
year         = {2026},
address      = {Washington, DC, USA},
howpublished = {\url{https://www.ferc.gov/rm26-4}},
note         = {Online; accessed 9 June 2026}
}

@article{miller2005peerprediction,
author  = {Miller, Nolan and Resnick, Paul and Zeckhauser, Richard},
title   = {Eliciting Informative Feedback: The Peer-Prediction Method},
journal = {Management Science},
volume  = {51},
number  = {9},
pages   = {1359--1373},
year    = {2005},
doi     = {10.1287/mnsc.1050.0379}
}

@misc{zhao2024peerprediction,
author        = {Zhao, Zishuo and Chen, Xi and Zhou, Yuan},
title         = {It Takes Two: A Peer-Prediction Solution for Blockchain Verifier's Dilemma},
year          = {2024},
eprint        = {2406.01794},
archivePrefix = {arXiv},
primaryClass  = {cs.GT}
}

@article{wasa2022contract,
author  = {Wasa, Yasuaki and Hirata, Kenji and Uchida, Kenko},
title   = {A Contract Theory Approach to Dynamic Incentive Mechanism and Control Synthesis for Moral Hazard in Power Grids},
journal = {IEEE Transactions on Control Systems Technology},
volume  = {30},
number  = {5},
pages   = {2072--2083},
year    = {2022},
doi     = {10.1109/TCST.2021.3130922}
}

@article{alasseur2020principalagent,
author  = {Alasseur, Cl{\'e}mence and Farhat, Heythem and Saguan, Marcelo},
title   = {A Principal-Agent Approach to Capacity Remuneration Mechanisms},
journal = {International Journal of Theoretical and Applied Finance},
volume  = {23},
number  = {08},
pages   = {2050052},
year    = {2020},
doi     = {10.1142/S0219024920500526}
}

@misc{barreto2014demandresponse,
author        = {Barreto, Carlos and Mojica-Nava, Eduardo and Quijano, Nicanor},
title         = {Incentives-Based Mechanism for Efficient Demand Response Programs},
year          = {2014},
eprint        = {1408.5366},
archivePrefix = {arXiv},
primaryClass  = {cs.SY}
}

@article{weidlich2008agent,
author  = {Weidlich, Anke and Veit, Daniel},
title   = {A Critical Survey of Agent-Based Wholesale Electricity Market Models},
journal = {Energy Economics},
volume  = {30},
number  = {4},
pages   = {1728--1759},
year    = {2008},
doi     = {10.1016/j.eneco.2008.01.003}
}

@techreport{pjm2025loadforecast,
  author      = {{PJM Interconnection}},
  title       = {2025 {PJM} Long-Term Load Forecast Report},
  institution = {PJM Interconnection},
  address     = {Audubon, PA, USA},
  year        = {2025},
  month       = jan,
  url         = {https://www.pjm.com/-/media/DotCom/library/reports-notices/load-forecast/2025-load-report.pdf},
  note        = {Posted January 24, 2025; accompanying forecast
                 tables include Table B-9b; accessed August 3, 2026}
}

\end{document}